\documentclass[useAMS,usenatbib]{mnras}
\usepackage{epsfig}
\usepackage{graphicx}
\usepackage{pdflscape}
\usepackage{textcomp}
\usepackage{float}
\usepackage{comment} 
\usepackage[caption = false]{subfig}

\newcommand{\ltsima} {$\; \buildrel < \over \sim \;$}
\newcommand{\gtsima} {$\; \buildrel > \over \sim \;$}
\newcommand{\lta} {\lower.5ex\hbox{\ltsima}}
\newcommand{\gta} {\lower.5ex\hbox{\gtsima}}
\newcommand{\Epk} {E_{\rm pk}}

\title[Clustering of GBM GRBs]{Clustering of gamma-ray burst types in the {\it Fermi}-GBM catalogue: indications of photosphere and synchrotron emissions during the prompt phase}
\author[Z. Acuner et al.]{Zeynep Acuner$^{1,2}$\thanks{email: acuner@kth.se} \&  
Felix Ryde$^{1,2}$\\
\\
$^{1}$Department of Physics, KTH Royal Institute of Technology, AlbaNova, SE-106 91 Stockholm, Sweden\\ 
$^{2}$The Oskar Klein Centre for Cosmoparticle Physics, AlbaNova, SE-106 91 Stockholm, Sweden\\ 
}

\begin{document}

\date{Accepted... Received...; in original form ...}

\pagerange{\pageref{firstpage}--\pageref{lastpage}} \pubyear{2017}

\maketitle

\label{firstpage}

\begin{abstract}
Many different physical processes have been suggested to explain the prompt gamma-ray emission in gamma-ray bursts (GRBs). Although there are examples of both bursts with photospheric and synchrotron emission origins, these distinct spectral appearances have not
been generalized to large samples of GRBs. Here, we search for signatures of the different emission mechanisms in the full {\it Fermi Gamma-ray Space Telescope}/GBM catalogue. We use Gaussian Mixture Models to cluster bursts according to their parameters from the Band function ($\alpha$, $\beta$, and $\Epk$) as well as their fluence and $T_{90}$. We find five distinct clusters. We further argue that these clusters can be divided into bursts of photospheric origin (2/3 of all bursts, divided into 3 clusters) and bursts of synchrotron origin (1/3 of all bursts, divided into 2 clusters). For instance, the cluster that contains predominantly short bursts is consistent of photospheric emission origin. We discuss several reasons that can determine which cluster a burst belongs to: jet dissipation pattern and/or the jet content, or viewing angle.

\end{abstract}


\section{Introduction}  
\label{sec:intro}
Gamma-ray bursts are holding one of the mysteries in high-energy astrophysics, currently evading a complete picture explaining the physics of their spectra. Still, significant progress has been made since their discovery including many attempts to describe the spectra in different physical frameworks. These include emission due to internal or external shocks, which are assumed to be non-thermal in nature ~\citep{Katz1994, Tavani1996, Rees1994, Sari1998} and emission from the photosphere as predicted to occur in the fireball model ~\citep{Meszaros&Rees2000, Rees&Meszaros2005, Peer2007, Thompson2007}. To assess the applicability of a physical model, typically the photon index, $\alpha$, of the sub-peak power-law in the spectrum is studied ~\citep[e.g.][]{Preece1998, Axelsson2015, Yu2015}. The distribution of $\alpha$ has a characteristic peak at a value around $\alpha \sim -0.85$ \citep{Burgess2014a}, which coincides with the value expected for slow-cooled synchrotron emission. Asymptotically this slope is $\alpha= -2/3$ \citep{Tavani1996}, however,
as \citet{Burgess2014a} pointed out  the asymptotic synchrotron slope is rarely reached,  due to the limited energy range of the fitted data. By simulating the observed spectra with a synchrotron model, they showed that one should not expect a very sharp peak around $\alpha= -2/3$, but the peak value is instead expected to be at $\alpha \sim -0.8$. Similarly, a dispersion of measured values is expected to give rise to a width of around $\Delta \alpha \sim 0.5$. The coincidence of the observed and expected peak of the $\alpha$-distribution has thus naturally been used as an argument for synchrotron emission. 
However, the synchrotron model is confronted by several issues. First, with typical physical assumptions the cooling is required to be fast rather than slow leading to an expected distribution peak at $\alpha \sim -1.5$.  Various non-trivial physical settings have been discussed to reconcile observations ~\citep[e.g.][]{Daigne&Mochkovitch2002, Uhm&Zhang2014, Beniamini&Piran2013}, but in all cases a substantial fraction of burst spectra are left unaccounted for. On the other hand, the photospheric model can account for a large diversity of spectra if subphotospheric dissipation ~\citep[e.g.][]{Rees&Meszaros2005, Peer2006, Giannios2006, Beloborodov2010, Vurm2011} and/or high latitude effects \citep{Lundman2013, ito2013}
are taken into account. However, the location of the peak in the $\alpha$-distribution needs to have an natural explanation. One such possibility was given by \citet{Lundman2013}, who argued that high-latitude emission from the photosphere can give $\alpha \sim - 1$, in the case of narrow jets with an opening angle of the order of the inverse of the Lorentz factor of the flow. However, most bursts are estimated to have broader jets (e.g. \citep{goldstein2016, LeMehta2017}). Moreover, \citet{Vurm2016} argued that $\alpha \sim -1$ is a natural consequence of unsaturated Comptonisation of soft synchrotron photons produced below the photosphere. However, it is unclear how burst with unpronounced peaks ($\alpha \sim \beta$) are formed in such a scenario.

It has therefore been suggested that there is an interplay between different emission mechanisms, either alone or combined with each other \citep[e.g.][]{Meszaros&Rees2000, Ryde2005, Battelino2007, Guiriec2011, Guiriec2013}.    
It is then a natural consequence that subgroups of GRBs could exist, that are produced by different emission mechanisms \citep[e.g.][]{BegueBurgess2016}.
Indeed, the observations of bursts with multiple components producing a mixture of thermal and non-thermal spectra ~\citep[][etc.]{Ryde2005,RydePeer2009, Guiriec2010, Axelsson2012, Guiriec2016, Nappo2017} further strengthens the case that large samples of GRBs are more likely to be explained by making use of several different physical emission mechanisms.

The hypothesis of separate physical sources of different groups of GRBs, implies that these groups should have different characteristics, when it comes to spectral shape, spectral components, variability and morphology of the light curves, and correlations between such characteristics. This fact motivates a search for possible statistical groupings of GRBs in large data samples.  Previously, several clustering studies of {\it CGRO} BATSE bursts have been performed \citep[e.g.][]{hakkila2003, horvath2006,  chattopadhyay2017}. These studies agree on the existence of three major groups of GRBs in which the main classification is based on fluence and $T_{\rm 90}$ measures and indicates that GRBs are divided in two, as short and long bursts, latter of which further divides into high (long $T_{\rm 90}$) and low fluence (intermediate $T_{\rm 90}$) classes.  In the present study, we search for clusters in the catalogue of bursts observed by the Gamma-ray burst monitor (GBM) onboard the {\it Fermi Gamma-ray Space Telescope} and further examine their spectral and temporal properties. We find that bursts can be classified as pre-dominantly thermal or non-thermal bursts, with clustering also strongly separating between long and short bursts. The outline of the paper is as follows: the method and data used for the clustering performed is explained in Section \ref{sec: 2}, the clustering analysis and results are reported in Section \ref{sec: 3}, the findings are discussed in Section \ref{sec: 4} and a general conclusion is derived in Section \ref{sec: 5}.

\section{Clustering analysis: data sample and method}    
\label{sec: 2}

\subsection{Data Sample}
This study makes use of the {\it Fermi} GBM burst catalogue published at HEASARC\footnote{www.heasarc.gsfc.gov/} which provides an extensive burst sample with their spectral properties and different model fit parameters. We use all available bursts observed until 14 February 2017, for which there are automatic spectral fits provided. We make use of the fits that are performed on the time-resolved data around the flux peak, within the time interval given in the GBM catalogue.  We select all the bursts for which a \citet{Band1993} function has successfully been fit. This includes removing 141 bursts for which the Band fit is not provided or for which the parameter errors are not determined. We note that these 141 unsuccessful fits all occurred before July 2012, and most likely are due to malfunctioning of the automated fitting algorithm. Moreover, the properties of these 141 bursts are similar to that of the entire catalogue. We, therefore, conclude that the omission of these bursts do not pose any selection bias to our study. The resulting sample consists of 1692 bursts.

The Band function is an empirical function that is traditionally used to describe GRB prompt spectra. It is a smoothly broken power law with four variables: the low energy power law index $\alpha$ (the photon flux $N_{\rm E} \propto E^{\alpha}$), the high energy power law index $\beta$, the energy of the spectral break\footnote{Note that originally the break was defined as the $e$-folding energy $E_{\rm 0}$, related as $E_{\rm pk} = (2+\alpha) E_{\rm 0}$ } in the $\nu F_\nu$ spectrum $E_{\rm pk}$, and the normalisation ~\citep{Band1993}. Even though the Band function is hugely successful in fitting and characterizing GRB spectra, we note that it is not the best fit for all spectra. 
The GBM catalogue in many cases report another model as the best fit model. In most cases this is a cut-off power-law model, which is similar to the Band function but does not have a high energy power-law. However, the selection of best fit model, that is made for the GBM catalogue, is based purely on the c-stat values. Such a decision is fast but not robust, since simulations are required to assess the statistical preference, which is different for each bursts \citep{Gruber2014}. Moreover, the difference in c-stat values between models are in most cases small. In addition to this, bursts have been reported to have extra spectral components, which are not tested for in the catalogue. These include power-law components \citep[][etc]{Gonzalez2003, Ryde2005,Abdo2009_nature, Ackermann2010}, several spectral breaks \citep[][etc.]{Barat1998, RydePeer2009, Iyyani2013}, high energy cut-offs \citep{Nava2011, Ahlgren15, Vianello2017}. Nevertheless, we consistently use the Band function fit for all bursts, since our purpose is mainly to  capture characteristic differences in spectral shapes. For this purpose the Band function fits are sufficient.

\begin{figure}
\includegraphics[width=\columnwidth, height=0.35\textheight]{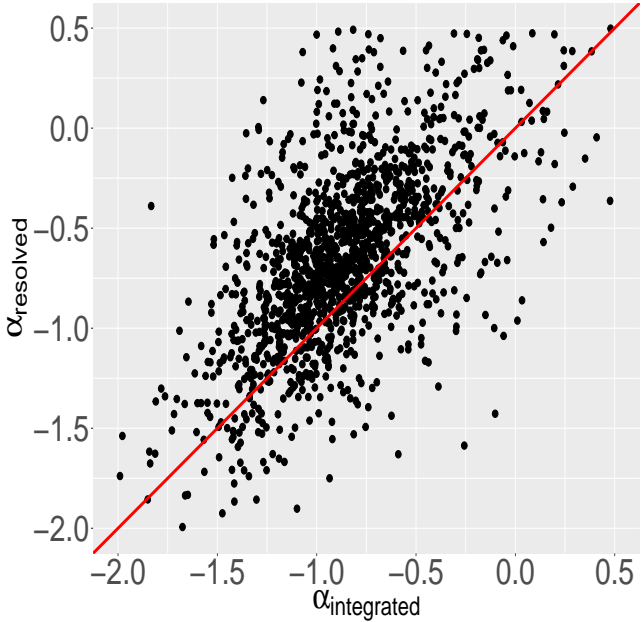}
\caption{Comparison of low-energy power-law index ($\alpha$) from the time-resolved (peak-flux) and time-integrated spectra (see also \citet{Nava2011}). For clarity we have made an $\alpha$-error cut of $\Delta \alpha < 0.5$. The medians of the uncertainties on $\alpha_{integrated}$ and  $\alpha_{resolved}$, in the figure, are 0.09 and 0.19 respectively}.
\label{fig:fig1}
\end{figure}

\begin{figure}
\includegraphics[width=\columnwidth, height=0.35\textheight]{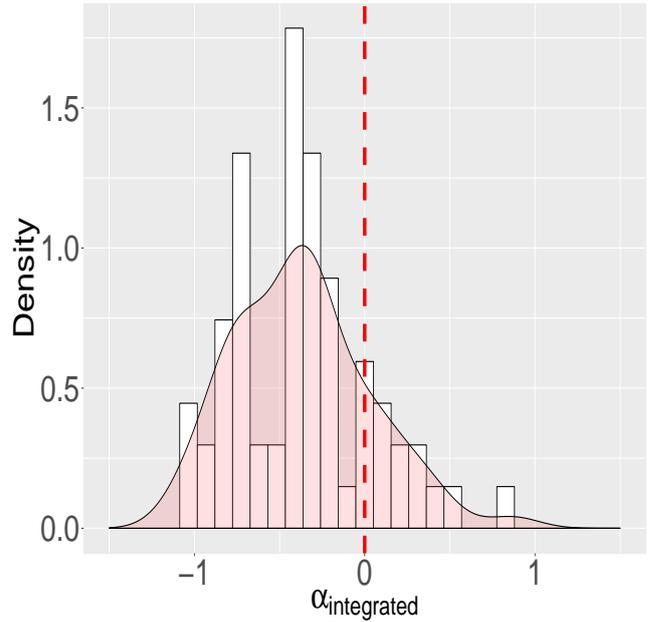}
\caption{Histogram for $\alpha_{integrated}$ with a cut for $\alpha_{resolved}$ greater than 0 in the GBM sample (for clarity $\Delta \alpha < 0.5$).}
\label{fig:fig2}
\end{figure}

The online catalogue provides spectral fits to the emission from around the peak of the light curve (time resolved spectrum) as well as fits to the emission from the entire duration (over their full fluence; time integrated spectrum). We exclusively study the time-resolved spectra, since they carry the cleanest signature of the underlying emission physics. The reason is that there typically is strong spectral evolution during bursts \citep{golenetski83, kargatis95} 
and therefore the integrated emission will be smeared out. To illustrate this point, we plot in Figure \ref{fig:fig1} the relation between the Band $\alpha$ parameter measured from the time-resolved spectrum versus $\alpha$ measured from the time-integrated spectrum (see also \citet{Nava2011}). The equal line is given by the red line. It is evident that the time integrated spectra are significantly softer than then time resolved spectra, since most of the points lie above the equal line. This means that time-integrated spectra cannot be used to directly assess the underlying emission physics, without assuming something about the spectral evolution and thereby smearing of the spectra \citep{ryde1999}.
Indeed, if we select only bursts for which $\alpha_{\rm resolved} > 0$, i.e.  bursts that can only be explained by thermal emission mechanisms, the corresponding  $\alpha_{\rm integrated}$ will have a broad distribution, many of which would falsely be considered consistent with synchrotron emission since the slope is $\alpha $ \textless $-2/3$, as shown in Figure \ref{fig:fig2}.

In addition to the spectral shape, described by the \citet{Band1993} function fits, we use the catalogued values for the bursts duration, measured by $T_{\rm 90}$, and the energy fluence. The fluence is the energy flux integrated over the duration, $T_{
\rm 90}$, of the burst. 
The reason we choose to use the fluence, and not the peak flux, is that for our purposes it is the  most appropriate measure to use \citep{PetrosianLee1996}. While the fluence corresponds to the total emitted energy of the GRB, the peak flux reflects the momentary variation of the variable flow, for instance, of the bulk Lorentz factor. More importantly, the peak flux value depends on the integration time, which typically is much larger than the intrinsic physical time scale.

In the analysis below, we will also make use of the spectral width around the $\nu F_\nu$ peak, that is determined for a fraction of these bursts \citep{Axelsson2015},  giving a sample of 689 bursts. In addition, we will use the time variability of the light curve, that has been determined for another sub-set of bursts \citep{Golkhou&Butler2014, Golkhou&Butler2015} consisting of 804 bursts (out of the total 938 bursts, we only used the ones that matched with our GBM catalog selection of 1151 bursts). The latter two groups of bursts were used for assessing and verifying the clusters obtained for the fundamental parameters from the full GBM catalogue GRBs.

\subsection{Method}
\label{sec:22}

\subsubsection{Data pre-processing}

We use the following parameters from the GBM catalogue  for the clustering searches:
$\alpha, \beta, E_{\rm pk}, T_{\rm 90}$, and fluence. 
The distributions of these variables are very skewed with $\beta$ being the most problematic, revealing a Rayleigh type distribution with a very long tail. This type of a distribution can cause the clustering results to be less precise. To remedy this issue, we performed several cuts on $\beta$ and examined the resulting quantile-quantile plots (QQ-plots). The cut at $\beta$ \textgreater $-4$ was sufficiently successful in removing the very heavy tail. This leaves 1151 bursts for the main sample in contrast to 1692 burst from the raw sample. We also separately study the sample with very steep $\beta$ ($\beta$ \textless $-4$).

The further examination of all variables suggested the need of a transformation that would Gaussianize the data by gathering the outliers closer to the mean. This is an important step since Gaussian data is more manageable, with intuitive mean and median results, furthermore, normality is a requirement for the clustering method Gaussian Mixture Models (GMM) used in this paper. To achieve this, we have made use of the Box-Cox transformation (\citet{boxcox}) implemented in the R package \texttt{forecast} \citep{forecast}. This method both quantifies the deviations from Gaussianity in the data as well as later takes this numerical quantity as an input for producing the transformed versions of the variables and hence is more tailored to the specific data set at hand compared to a plain logarithmic transformation. Since neither logarithmic nor Box-Cox transformation can deal with negative data sets, appropriate constants were added to the variables with negative values before carrying out this step. The resulting transformed variables were examined with QQ-plots once more, to identify any significant outliers that may distort the clustering. Following this, the sample was scaled and centered by making use of the R function \texttt{scale} \citep{R} which includes subtracting the mean of the parameter from every element and diving them by the standard deviation. This is carried out so that the features which have broader range of values do not dominate the overall variability in the data. 

Before feeding the data into different clustering methods, we have performed a principal component analysis (PCA) via the R package \texttt{factoMineR} \citep{factominer} to be able to reduce our highly dimensional multivariate data set to a lower-dimensional set. This allows being able to select the most dominant components in the data, while removing strong inherent correlations that might affect the clustering results. 

\subsubsection{Clustering}

After the pre-processing, the data was fed into the Gaussian Mixture Model implementation \texttt{mClust} \citep{mclust} in R. GMM provided a feasible background of information for interpretation of the results since it is model based and hence can give probabilities for each burst belonging to each group. The clustering was performed with selected non-parametric and density based clustering algorithms as well and the results were found compatible with those given by GMM. The number of clusters were determined by the Bayesian Information Criterion (BIC) implemented in the Expectation Maximization (EM) algorithm in \texttt{mClust} which assessed the optimal number of clusters. Output of this method was 5 GMM clusters with different cluster sizes. 

The resulting clusters were evaluated by calculating the Silhouette scores with the method \texttt{silhouette} \citep{silhouette} in the package \texttt{cluster} \citep{cluster}. With this method, we were able to single out the bursts that were assigned to the wrong clusters during the initial GMM clustering. By re-assigning these bursts to their correct clusters, we were further able to improve the Silhoutte scores and the precision of the clusters at hand.

\subsubsection{Spectral analysis of representative bursts}
\label{sec:spectral}
Understanding the spectral morphological specifications of the different classes of bursts that are revealed by the clustering procedure is the primary goal of this study. The clustering analysis is based on the standard spectral fits by using the Band function. However, the actual spectrum might not be best described by a Band function. 

For understanding the details of each cluster, we have created  shortlists which are presented in Appendix B with a selection criterion that maximizes the probability of belonging to a group while minimizing the error on variable $\alpha$ to ensure good convergence of the Band fit in the catalog values. The top bursts from these shortlists were analysed spectrally.

To carry out this task, we work with the data from the Gamma-ray Burst Monitor (GBM;~\cite{Meegan2009}
) on board the \textit{Fermi Gamma-ray Space Telescope}. GBM harbors 12 sodium iodide (NaI) detectors that observe between 8 keV and 1 MeV as well as two Bismuth Germanate (BGO) detectors which are sensitive to a higher energy range of {\bf 200} keV to 40 MeV. We use the time-tagged event data (\texttt{tte}) and the standard response files as provided by the GBM team. We use the source and background intervals given in the GBM catalogue for consistency with the Band fit parameters. For the analysis, we use \texttt{Xspec} spectral fitting package \citep{Arnaud1996} and we produce the PHA files to be used in \texttt{Xspec} via \texttt{gtBurst} \citep{Vianello}. 
To be able to assess a large range of different spectral shapes, we use 8 different empirical spectral models: (i) the Band function, (ii) the Band function and a powerlaw, (iii) the Band function and a blackbody, (iv) the \textit{compTT} model, which is an analytic model that describes the Comptonization of soft photons in a hot plasma \citep{Titarchuk}, (v) a doubly broken powerlaw model with a parameter to adjust for the smoothness of the breaks (defined in \ref{sec:6}), (vi) a single blackbody, (vii) a blackbody and a powerlaw, (viii) two blackbodies and finally, (ix) two blackbodies and a powerlaw. With these models, we were able to probe the number of breaks in each spectra, as well as the existence of an additional powerlaw component. We point out that these models are used to empirically assess the shape of the spectra and a thorough physical modelling is later needed for the interpretation of the radiation mechanisms, such is done in \citet{Iyyani2015, Ahlgren15,Vurm2011, Vianello2017}.

\section{Clustering results of main sample $\beta > -4$}
\label{sec: 3}
\begin{figure}
\includegraphics[width=\columnwidth, height=0.35\textheight]{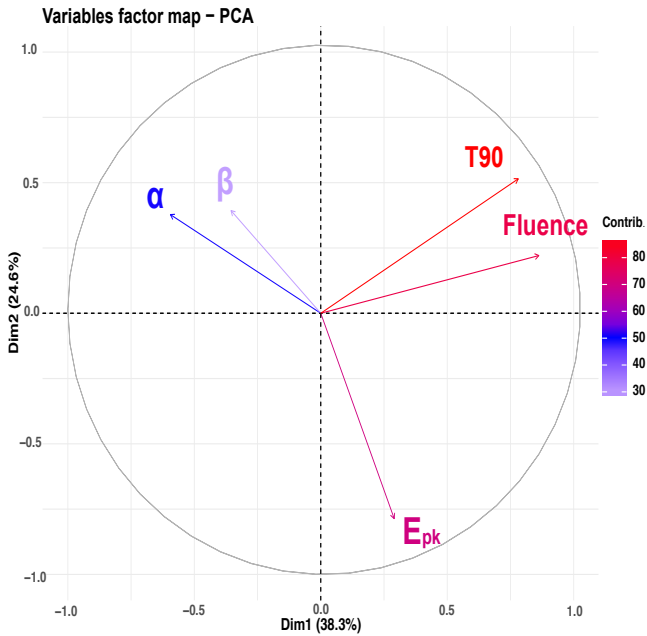}
\caption{Contributions of the five variables to first (x-axis) and second (y-axis) PCA dimensions represented inside the circle of correlations. The percentage of total contribution is given by the color coding and the angles to the two axes are indicative of the percentage of contributions to either dimension.}
\label{fig:figpca}
\end{figure}

The PCA analysis shows that the main variability in the data set is from $E_{pk}$, fluence and $T_{\rm 90}$ which drive the clustering (\autoref{fig:figpca}). The selection of two primary PCA components for clustering, which explain 63 per cent of the variance in the data, neatly reduces the number of clusters to three which is in accordance with the main groups of fluence and $T_{\rm 90}$ that were described in Section \ref{sec:intro}. However, since this study strives for an explanation of spectral morphology, we use all 5 components for the clustering to make use of the more minor variability in parameters $\alpha$ and $\beta$ which results in 5 GMM clusters. The final result of the clustering analysis is given in Table \ref{tab:1} with values of variables used in the clustering for each cluster. These properties are clearly distinguished between the groups and the most notable are the large $E_{\rm pk}$ values in clusters 2 and 5 and the short durations in cluster 5.

\begin{table*}
	
    \centering
	\caption{The list of samples sizes, means, standard deviations (SD), medians and inter-quantile ranges (IQR) for the five variables used in the clustering for five GMM clusters.}
	\label{tab:1}
    \begin{tabular}{lllllllllllllllllllll} 
		\hline
        \hline
      Cluster ID   & Sample size& Band parameters (mean(SD),  median,                 IQR) & ~&   ~ &    ~ &  ~ &  ~&   ~                  &  ~ &  ~&~ &~       &  ~ &  ~&   ~ &   ~    &   ~ &   ~   \\ \hline
    \end{tabular}
    \begin{tabular}{lllllllll}
    ~       & ~      &~  &~    & $E_{pk}$ [keV]  & $\alpha$ & $\beta$ & Fluence [erg/cm\textsuperscript{2}]         & $T_{90} $[s]                    \\   \hline
    Cluster 1 & ~  &369   &~      & 139(85), 206, 86              &  -0.36(0.67), -0.48, 0.6            & -2.9(0.4), -2.8, 0.7            & 7(9), 4, 6$ \times 10^{-6}$  & 31(39), 19, 27                          \\ \hline
    Cluster 2 & ~  &381   &~        & 503(616), 645, 345              &  -0.74(0.31), -0.77, 0.4               & -2.4(0.4), -2.3, 0.5              & 3(6), 1.5, 3$ \times 10^{-5}$  & 76(94), 47, 71                            \\ \hline
    Cluster 3 &~  & 233  &~        & 94(72), 118, 79               & 0.44(1.42), -0.05, 1.5              & -1.9(0.2), -1.9, 0.3              & 3(3), 2, 2$ \times 10^{-6}$  & 35(49), 24, 35                             \\ \hline
    Cluster 4 & ~  &40   &~      & 242(377), 323, 141               & -1.47(0.14), -1.43, 0.2              & -2.4(0.6), -2.2, 0.9             & 5(7), 2, 6$\times10^{-6}$ & 36(79), 37, 51                            \\ \hline
    Cluster 5 & ~  &128  &~       & 604(664), 144, 588             & 0.7(2.82), -0.09, 1.2              & -2.3(0.6), -2.2, 0.8             & 8(0.1), 4, 5$\times 10^{-7}$ & 1.1(1.4), 0.54, 0.06                          \\ 
		\hline
	\end{tabular}
\end{table*} 

\subsection{Appearances in 2D plots}
\label{sec:2D}

\begin{figure}
\includegraphics[width=\columnwidth]{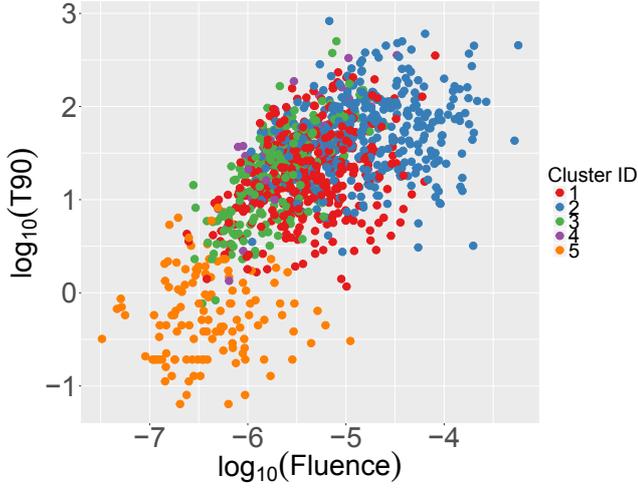}
\caption{Plot of log($T_{\rm 90}$) versus log(fluence) with color coding representing the five GMM clusters. The medians of the uncertainties on $T_{\rm 90}$ and fluence are 1.9 seconds and 5$\times 10^{-8}$ $erg/cm\textsuperscript{2}$ respectively.}
\label{fig:FT90}
\end{figure}

Before examining each cluster in more detail, it is interesting to asses them in different parameter spaces (Figures \ref{fig:FT90} to \ref{fig:betaf}) selected for their representative powers compared to other pairings. Note that the primary focus of this study is not to study correlations between observed quantities, but rather to identify classes of burst with different properties. Nevertheless, the two-dimensional plots are useful in illustrating where the different clusters dominate.

Figure \ref{fig:FT90} presents the GMM clusters in log($T_{90}$) versus log(Fluence). Apart from the distinct cluster 5 (low fluence, small $T_{\rm 90}$; orange in Fig. \ref{fig:FT90}) there is only a very weak correlation (coefficient of determination $R \sim 0.4$; correlation coefficient $\sim 0.64$). However, the remaining clusters do occupy different regions distinguished by fluence. In particular, we note that cluster 2 (blue) distinguishes itself from the low fluence clusters (green, red, purple). The Kolmogorov-Smirnov test rejects the hypothesis of these two groups being the from the same distribution with a $p-$value smaller than $2.2\times 10^{-16}$.

\begin{figure}
\includegraphics[width=\columnwidth]{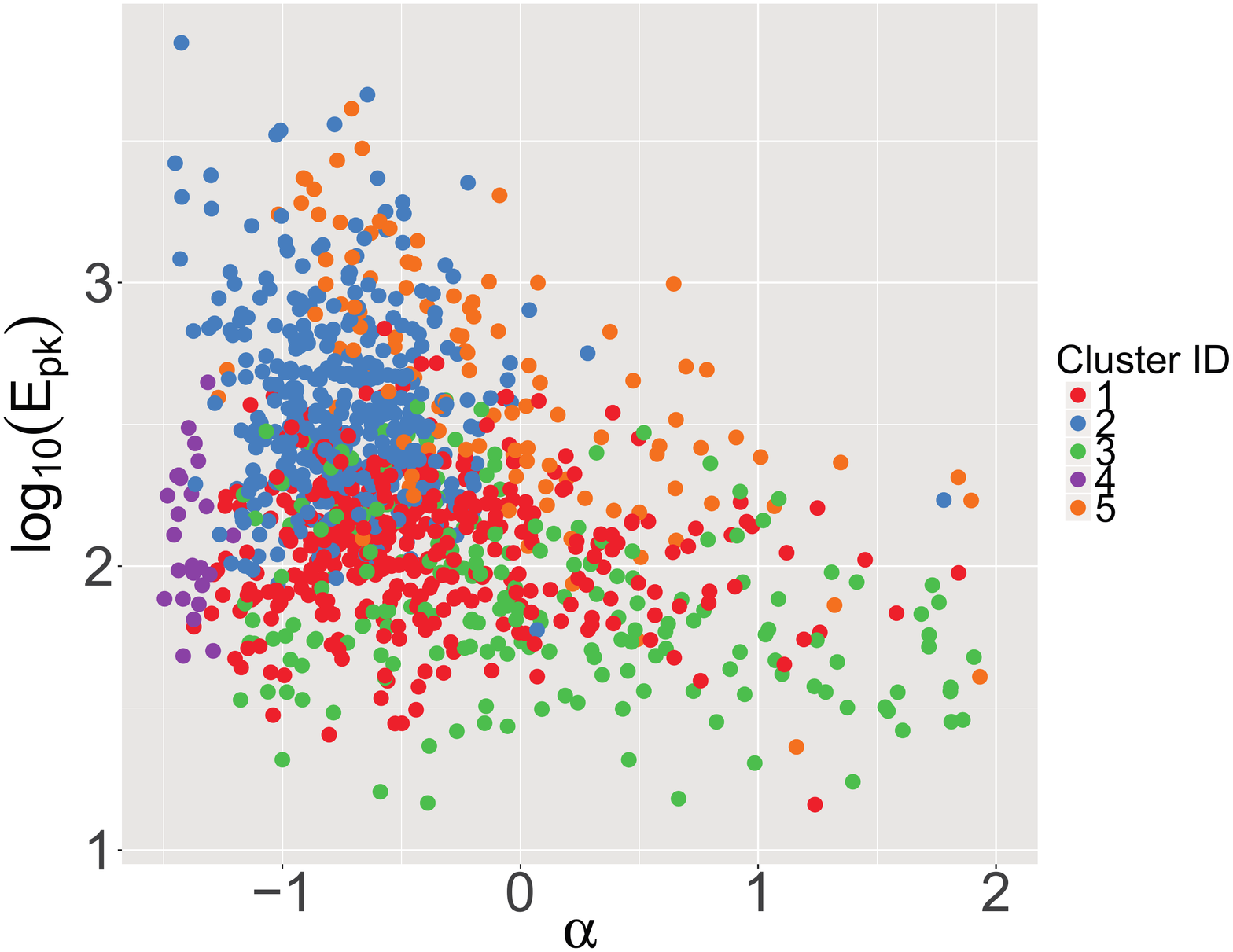}
\caption{Plot of log($\Epk$) versus $\alpha$ with color coding representing the five GMM clusters. The medians of the uncertainties on $\Epk$ and $\alpha$ are 46.6 keV and 0.3 respectively.}
\label{fig:Epka}
\end{figure}

Figure \ref{fig:Epka} presents the GMM clusters in log($E_{\rm pk}$) versus $\alpha$. For plotting purposes alone, we cut the $\alpha$-axis at $\alpha = +2$. The reason is that, first, this is the hardest slope expected theoretically since it is the sub-peak slope of a Wien spectrum. Second, all bursts with a measured value of $\alpha>2$ (46 bursts), have large measurement errors and  all are in fact consistent with the Rayleigh-Jeans' slope of $\alpha = 1$, to within one sigma. Note also that the GBM detector has a limited energy range, which imposes a restriction in the range of $E_{\rm pk}$. In the figure, one can see a division inside the group of long bursts from $\Epk \approx 400$ keV, where the clusters 2 (blue) remains at the high $\Epk$ side while clusters 1 (red) and 3 (green) have lower $\Epk$ values. The short bursts, cluster 5 (orange), have high overall $\Epk$ values around the same range with cluster 2 while cluster 4 (purple) has a intermediate range of $\Epk$s.

\begin{figure}
\includegraphics[width=\columnwidth]{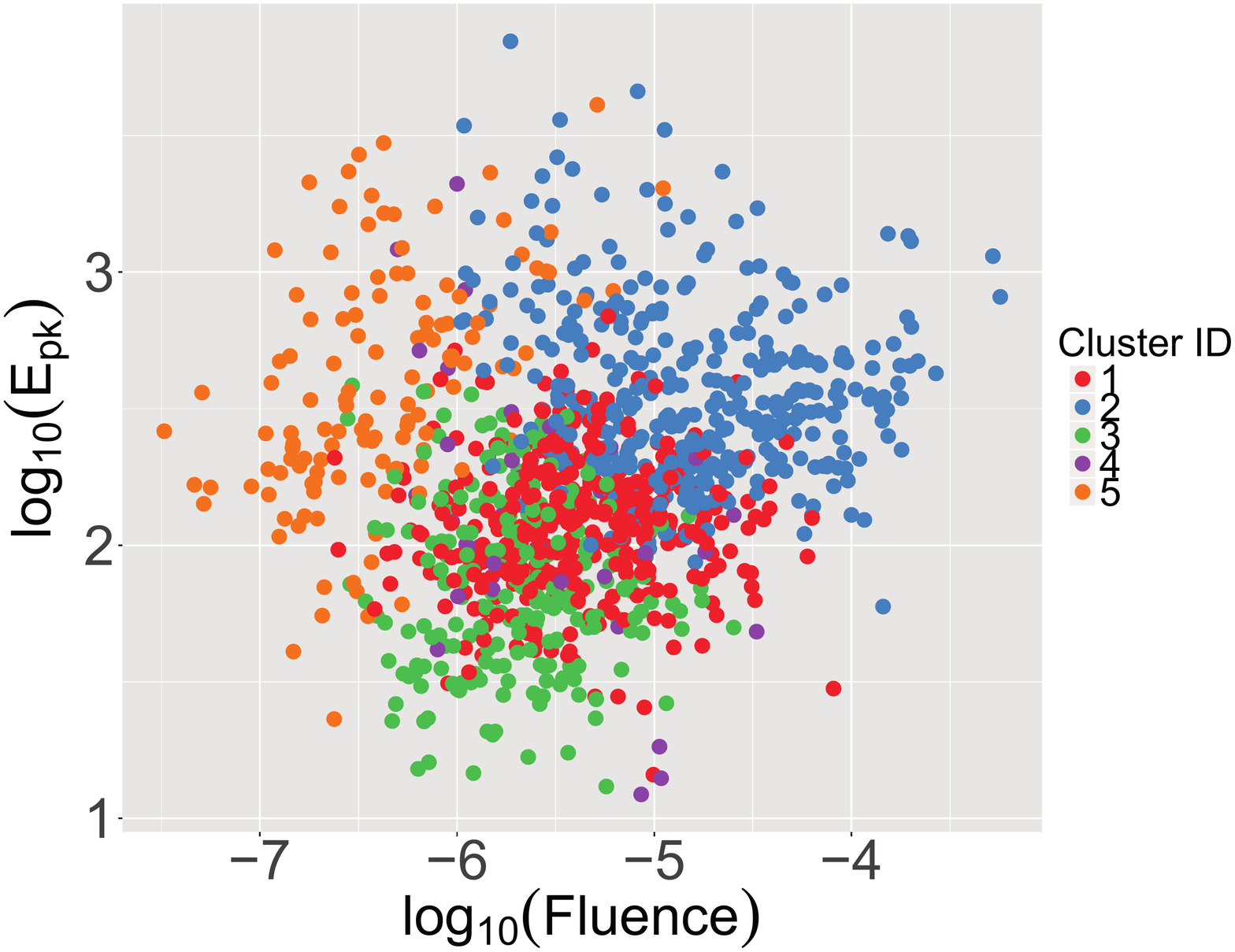}
\caption{Plot of log($\Epk$) versus log(fluence)  with color coding representing the five GMM clusters. The medians of the uncertainties on $\Epk$ and fluence are 46.6 keV and 5$\times 10^{-8}$ $erg/cm\textsuperscript{2}$ respectively.}
\label{fig:FEpk}
\end{figure}

Figure \ref{fig:FEpk} presents the GMM clusters in log($\Epk$) versus log(Fluence).  Cluster 2 is gathered in the region of high fluences and high peak energies, while the other extreme, cluster 3, is gathered at low $\Epk$ values and low fluences. The rest of the sample is gathered at intermediate fluence values and intermediate $\Epk$ values. The exception is the short burst cluster 5, which has very low fluences but comparably high $\Epk$s. This plot can be compared to figure 1 in Nava et al. (2008), which however only includes long burst, that is, cluster 5 does not appear in their figure.

\begin{figure}
\includegraphics[width=\columnwidth]{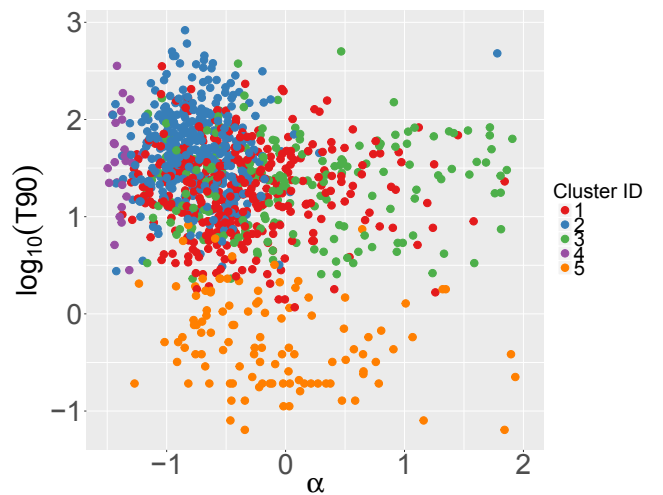}
\caption{Plot of log($T_{\rm 90}$) versus $\alpha$ with color coding representing the five GMM clusters. The medians of the uncertainties on $T_{\rm 90}$ and $\alpha$ are 1.9 seconds and 0.3 respectively.}
\label{fig:T90a}
\end{figure}

In Figure \ref{fig:T90a}, we plot log($T_{90}$) and $\alpha$ values. Cluster 5 is strikingly apart from the rest of the clusters thanks to its very short $T_{90}$ average. It is seen that clusters 1 and 3 have shorter $T_{\rm 90}$s and harder $\alpha$s compared to the longer bursts (clusters 2 and 4).

\begin{figure}
\includegraphics[width=\columnwidth]{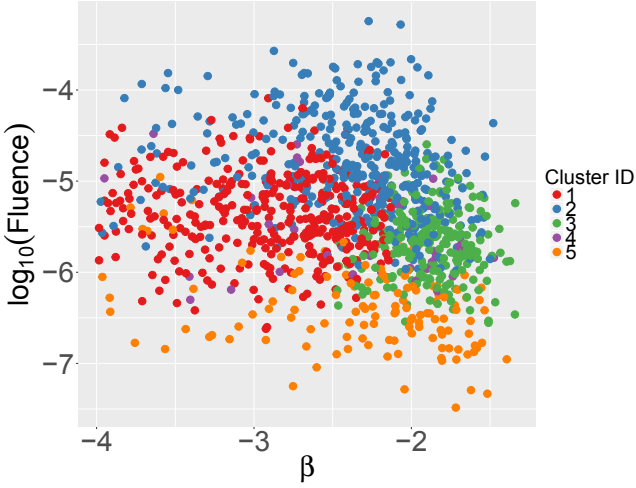}
\caption{Plot of log(fluence) versus $\beta$ with color coding representing the five GMM clusters. The medians of the uncertainties on $T_{\rm 90}$ and fluence are 5$\times 10^{-8}$ $erg/cm\textsuperscript{2}$ and 0.4 respectively.}
\label{fig:betae}
\end{figure}

Figures \ref{fig:betae} and \ref{fig:betaf} depict the relationship of $\beta$ to $\Epk$ and fluence, respectively. It is seen that the short burst cluster has the lowest fluences and highest (together with cluster 2) $\Epk$s as well as cluster 3 is very localized in the $\beta$ range with always lower $\Epk$ and fluence values than the rest of the long burst clusters.

Figure \ref{fig:fig10} displays the same parameters for Figures \ref{fig:Epka} and \ref{fig:T90a} but for the 50 bursts that belong to their clusters with highest probabilities assigned by the GMM. This provides a more concise view of how the clusters are separated.

\begin{figure}
\includegraphics[width=\columnwidth]{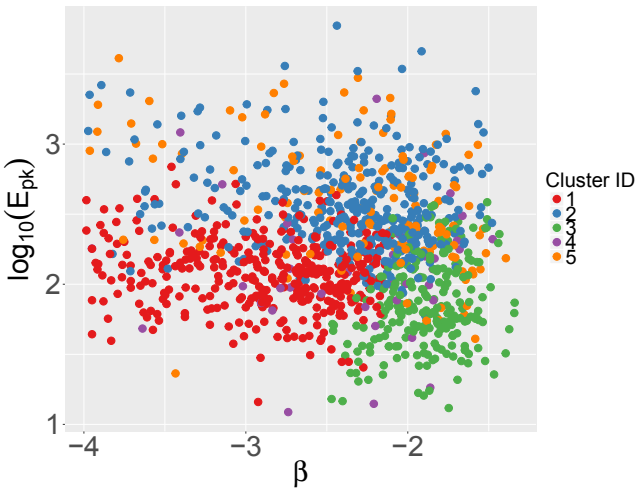}
\caption{Plot of log($\Epk$) versus $\beta$ with color coding representing the five GMM clusters.The medians of the uncertainties on $\Epk$ and $\beta$ are 46.6 keV and 0.4 respectively.}
\label{fig:betaf}
\end{figure}

\begin{figure}
\includegraphics[width=\columnwidth]{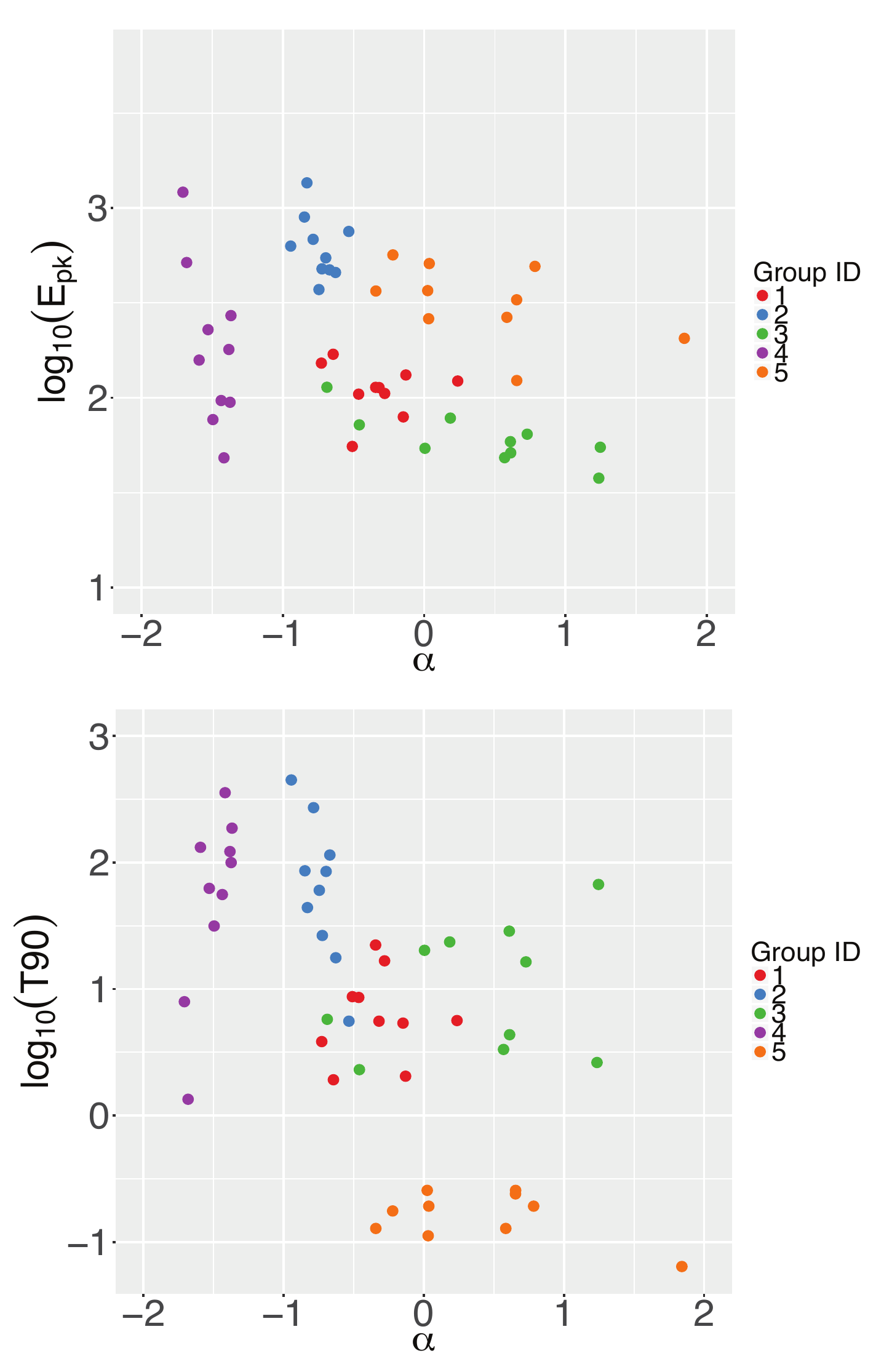}
\caption{Plots of log($\Epk$) versus $\alpha$ (cf. Fig. \ref{fig:Epka}) and log($T_{\rm 90}$) versus $\alpha$ (cf. Fig. \ref{fig:T90a}) for only the top 10 most probable bursts within each cluster, with color coding representing the five GMM clusters. The clusters now appear more distinctly.}
\label{fig:betaf}
\end{figure}

\subsection{Properties of bursts in the five clusters}

Based on the results in Table \ref{tab:1}, and alternatively in the figures above, the property character of the clusters can be identified.

As mentioned in the previous section, there exists two main categories defined by $T_{\rm 90}$ and separates the long ($T_{\rm 90}$$\textgreater$2) and short bursts ($T_{\rm 90}$$\textless$2) that have long been acknowledged as two distinct classes of bursts \citep{kouveliotou1993}. 
Furthermore, we detect the existence of two different type of long bursts that are separated as high and low fluence bursts (\S \ref{sec:2D}). These results are in accordance with previous studies that were focused on \texttt{BATSE} data (see \citet{} and \S \ref{sec:intro}). With the addition of the spectral parameters into the parameter space, we obtained a novel classification that combines the intrinsic and extrinsic properties of GRBs with clusters that both describe the spectral morphologies as well as the fluence and overall length of the bursts.

Now we are in a position to outline a general description of the burst sample at hand. There exists five clusters: 

(i) long bursts with low fluence values and narrow spectra consisting of short minimum variability time scales (cluster 1), 

(ii) very long bursts with high fluence values that consists of very short minimum variability time scales (cluster 2), 

(iii) intermediate length long bursts with very broad spectra and very long minimum variability time scales (cluster 3), 

(iv) intermediate length long bursts with broad spectra and long minimum variability time scales (cluster 4), 

(v) short bursts with very low fluence that have very short minimum time scale variabilities which can fundamentally be described by a single narrow component (cluster 5).

To further illustrate the differences between the clusters  we plot in Figure \ref{fig:violin} the $\alpha$-distributions as violin plots. Even though the $\alpha$-parameter only contains minor variability in the data set (\S \ref{sec:3}), and therefore is not dominant in forming the clusters, a clear distinction in $\alpha$-distributions is apparent. In particular, while clusters 2 and 4 are heavily occupying the $\alpha<0$ region,
clusters 1, 3 and 5 span the region between $\alpha = 0$ to $\alpha = 1$ as well, where only photospheric emission can reside (see further discussion in \S \ref{sec:51}).

In Section \ref{sec:31}, we discuss the properties of each group in more detail in the light of the parameters summarized in Tables \ref{tab:1} and \ref{tab:2}. In Section \ref{sec:32}, we discuss the spectral morphologies of each group independent of the Band fit parameters from the GBM catalog that were used to cluster the sample. Finally, in Section \ref{sec:33}, we propose a parameter to probe the temporal characteristics of GRBs which is later used to narrate the temporal morphology of each group.

\subsection{Characteristics of the clusters in terms of  Band parameters}  
\label{sec:31}

\subsubsection{Cluster 1}
Cluster 1 has an average $\alpha$ value of -0.36, which suggests a mediocre low energy powerlaw indice compared to other clusters. The sample size is quite large with 369 bursts and this cluster has the second highest fluence average among the clusters while having an $E_{pk}$ average of $\approx$ 140 keV which is the second lowest observed. $T_{\rm 90}$ averages approximately around 30 seconds  with the minimum time scale variability being around 1 second.

\subsubsection{Cluster 2}
Cluster 2 has the largest sample size of all groups with 381 bursts. This cluster has the second highest average $E_{pk}$ with $\approx$ 500 keV and a soft average low energy index, $\alpha$ with $\approx$ -0.7. It has the highest fluence average among our clusters. $T_{\rm 90}$ averages approximately at 70 seconds that indicates the longest bursts are gathered in this cluster which show a quite high variability in time with an average $\Delta t_{min}$ of $\approx$ 0.3 seconds. 

\subsubsection{Cluster 3}
Cluster 3 has the lowest average $E_{\rm pk}$ value at $\approx$ 90 keV as well as the second lowest average fluences with an $\alpha$ average of 0.5. This cluster has an intermediate $T_{\rm 90}$ average which approximates to 35 seconds. The average $\Delta t_{min}$ for Cluster 3 is $\approx$ 2 seconds, which makes the bursts in this group one of the least variable ones among the GBM cluster sample. This group is moderately populated with 233 members.

\subsubsection{Cluster 4}
The 4th cluster is the least populated cluster in our sample with 40 bursts. This cluster has an average $E_{pk}$ of $\approx$ 240 keV with a very soft $\alpha$ average ($\approx$ -1.5). The fluence and $T_{\rm 90}$ ($\approx$ ave. 40 seconds) are quite moderate with a  $\Delta t_{min}$ of $\approx$ 2 seconds.

\subsubsection{Cluster 5}
Cluster 5 has the highest $E_{\rm pk}$ average at $\approx$ 600 keV as well as the hardest average $\alpha$ value of $\approx$ 0.7 which suggests that this cluster is mainly dominated by bursts with very thermal spectra. However, the median $\alpha$ of the cluster is $\approx$ 0 so about half of the bursts exhibit negative low-energy indices. The cluster has the lowest fluence average which is mainly a consequence of its very low $T_{\rm 90}$ average of $\approx$1 seconds. Due to the strikingly short $T_{\rm 90}$ values of the bursts occupying this cluster, we identify Cluster 5 as consisting of the main majority of short bursts in our GBM sample. Indeed, while Clusters 1 to 4 have a total of 10 bursts that have $T_{\rm 90}$ $\textless$ 2, Cluster 5 has 121 bursts (out of a total of 128) that fall into the short burst category. Another property that distinguishes this cluster from the others is its highly variable lightcurve, with a minimum variability time scale average of $\approx$ 0.2 seconds which is the lowest among all clusters.

\begin{figure}
\includegraphics[width=1.2\columnwidth , keepaspectratio]{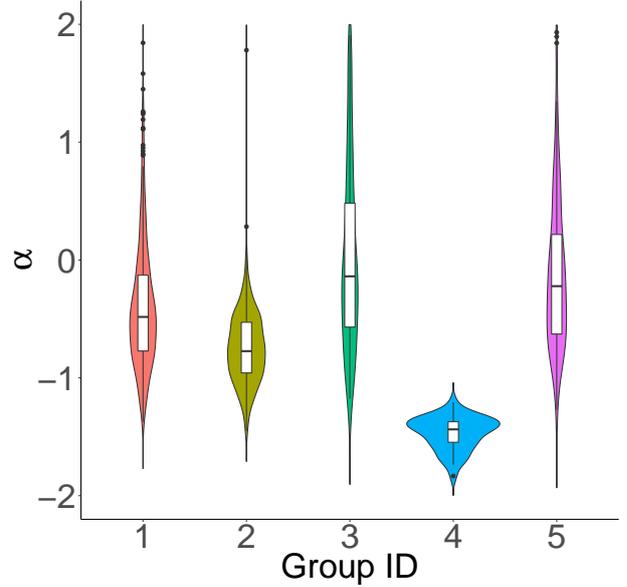}
\caption{Distribution of $\alpha$-values for the five clusters shown as violin plots. }
\label{fig:violin}
\end{figure}

\subsection{Clustering results of main sample $\beta < -4$}
\label{sec:comple}

The methods used in the clustering analysis above, require the parameter distributions to be Gaussian. Since the $\beta$ distribution is highly skewed we had to make a cut at a large value of $\beta =- 4$. We note that values less than $-4$ reproduces spectra that remain relatively similar, and all these bursts are close to having a exponential cutoffs.
 Nevertheless since the sample with $\beta < -4$ contains 541 bursts of the initial 1692 in our studied sample, we analyze it separately and compare the results with that of the main sample.

From this sample, six clusters were extracted following the method of Section \ref{sec:22}. Clusters in this sample emerge as the branches of the classes labeled as single break in the main sample (clusters 2 and 4 mostly). All 6 clusters have very soft $\alpha$ means with the softest being $\approx$ -1.7, belonging to a cluster that contains characteristically very broad spectra with high-energy, exponential cut-offs. The short bursts are again picked up by the clustering method in a single cluster, however, these have $\Epk$ values a few times higher than that of cluster 5 of the main sample. $T_{90}$, $\Epk$, and to a lesser extent $\alpha$, seem to be the major drivers of the variance in this sample.

\label{sec:3}

\section{Further analysis of cluster properties}

In this section, we further examine the properties of the bursts in the different clusters. 

Characteristics of the clusters which were not used in the clustering but nevertheless are helpful in interpreting the cluster are presented in two tables. Table \ref{tab:2} includes
the minimum variability time scale, $\Delta t_{min}$, the smoothness parameter, $S$, defined in eq. (\ref{eq:1}), and spectral width, while Table \ref{tab:3} includes the redshift, isotropic energy ($E_{\rm iso}$), and the peak flux at 64 ms.

\begin{table*}
	
    \centering
	\caption{The list of means, standard deviations (SD), medians and inter-quantile ranges (IQR) for three external parameters that are used in the interpretation of the five GMM clusters. Here $\Delta t_{min}$ is the minimum time scale variability, the smoothness parameter $S$ is defined in eq. \ref{eq:1}, and spectral width.}
	\label{tab:2}
    \begin{tabular}{llllllllllllllllllllllll} 
    \hline
    \hline
    Cluster ID     &~& External parameters (mean(SD), median, IQR) & ~&   ~ &    ~ &  ~      &  ~                               \\
    \hline
    \end{tabular}
    \begin{tabular}{llllll}
    ~ & ~  &      $\Delta t_{min}$ [s]                         & S                                & Width                             \\
    \hline
    Cluster 1  & ~& 				0.98(1.27), 0.48, 1.08         & 0.2(0.1), 0.2, 0.1    & 1.04(0.35), 1.12, 0.4   \\
    Cluster 2  & ~&  		       0.86(1.65), 0.34, 0.78    & 0.02(0.07), 0.02, 0.06 & 1.7(0.9), 1.3, 0.9   \\
    Cluster 3  & ~&                      2.02(3.68), 0.83, 1.98        & 0.1(0.2), 0.1, 0.2    & 2.8(1.4), 2.5, 2 \\
    Cluster 4  & ~&	       				 1.71(1.66), 1.15, 1.61         & 0.08(0.09), 0.05, 0.2  & 2.3(1.4), 1.6, 1.1 \\
    Cluster 5  & ~&	       				 0.16(0.42), 0.03, 0.06  & 0.2(0.2), 0.08, 0.2   & 1.29(0.78), 1.02, 0.6   \\
    \hline
    \end{tabular}
\end{table*}

\begin{table*}
	
    \centering
	\caption{The list of means, standard deviations (SD), medians and inter-quantile ranges (IQR) for three burst parameters that are used in the interpretation of the five GMM clusters. "NA" symbol stands for cluster parameters for which no $E_{iso}$ calculation was possible due to a lack of adequate number of bursts with measured redshift.}
	\label{tab:3}
    \begin{tabular}{llllllllllllllllllllllll} 
    \hline
    \hline
    Cluster ID     &~& External parameters (mean(SD), median, IQR) & ~&   ~ &    ~ &  ~ &  ~                                     \\
    \hline
    \end{tabular}
    \begin{tabular}{llllllll}
    ~ & ~  &     Redshift     &$E_{iso}$ [erg]		&Peak flux [pht/cm$^2$.s]                            \\
    \hline
    Cluster 1  & ~& 1.68(0.92), 1.69, 1.46 	&3(5), 0.9, 2$ \times10^{53}$	&15(22), 5, 11 						  \\
    Cluster 2  & ~&  	 1.57(0.82), 1.64, 0.8     &3(4), 1, 5$ \times10^{53}$		&27(67), 27, 17 				  \\
    Cluster 3  & ~&    2.48(1.21), 2.1, 1.92       &2(3), 0.5, 0.4$ \times10^{53}$		&3(4), 5, 3   			 \\
    Cluster 4  & ~&	   0.71(NA), 0.71, NA        &3(NA), 3, NA$ \times10^{52}$	&7(5), 6, 2   			 \\
    Cluster 5  & ~&	   NA(NA), NA, NA            &NA, NA, NA 	&13(18), 8, 8 						  \\
    \hline
    \end{tabular}
\end{table*}

\subsection{Spectral morphology of each cluster}
\label{sec:32}

In this section, we perform detailed spectral analysis on the bursts that belong to their respective clusters with the highest probabilities, as described in \S \ref{sec:spectral}. 

\begin{figure}
\includegraphics[angle=270, width=\columnwidth]{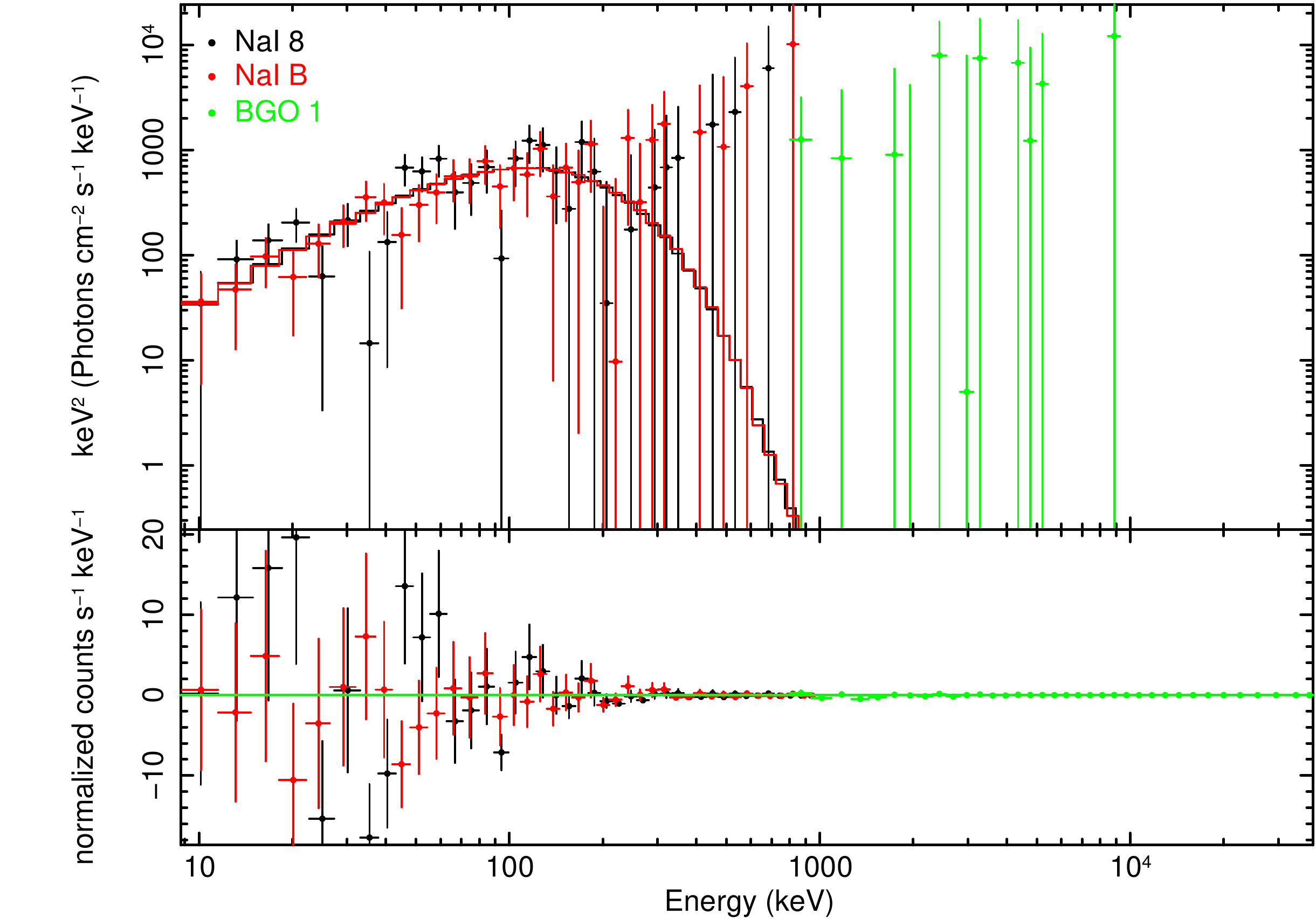}
\caption{Cluster 1 template burst, GRB100816024, fitted with the Band function.}
\label{fig:g1}
\end{figure}

\begin{figure}
\includegraphics[angle=270, width=\columnwidth]{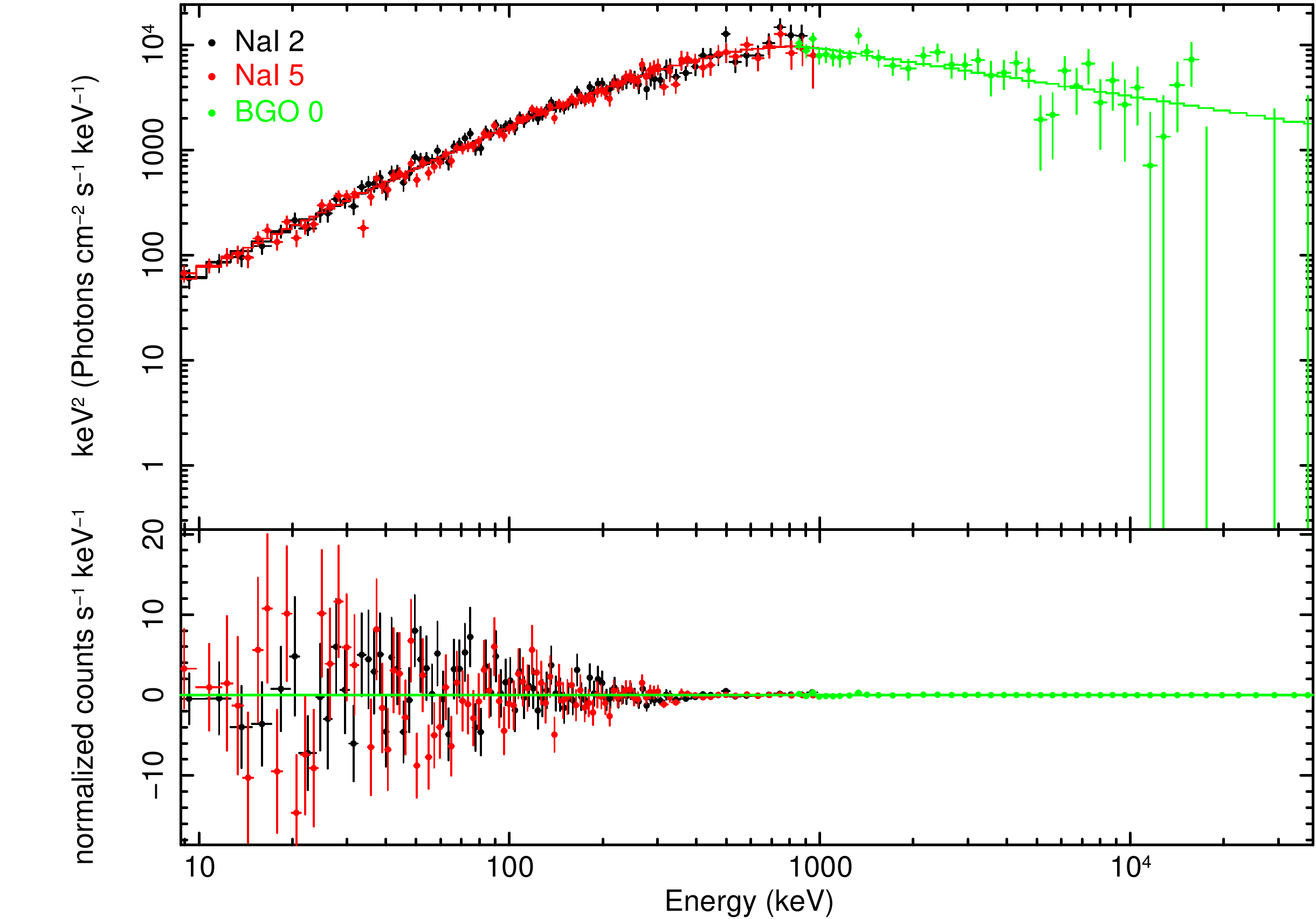}
\caption{Cluster 2 template burst, GRB12071115, fitted with the Band function.}
\label{fig:g2}
\end{figure}

\begin{figure}
\includegraphics[angle=270, width=\columnwidth]{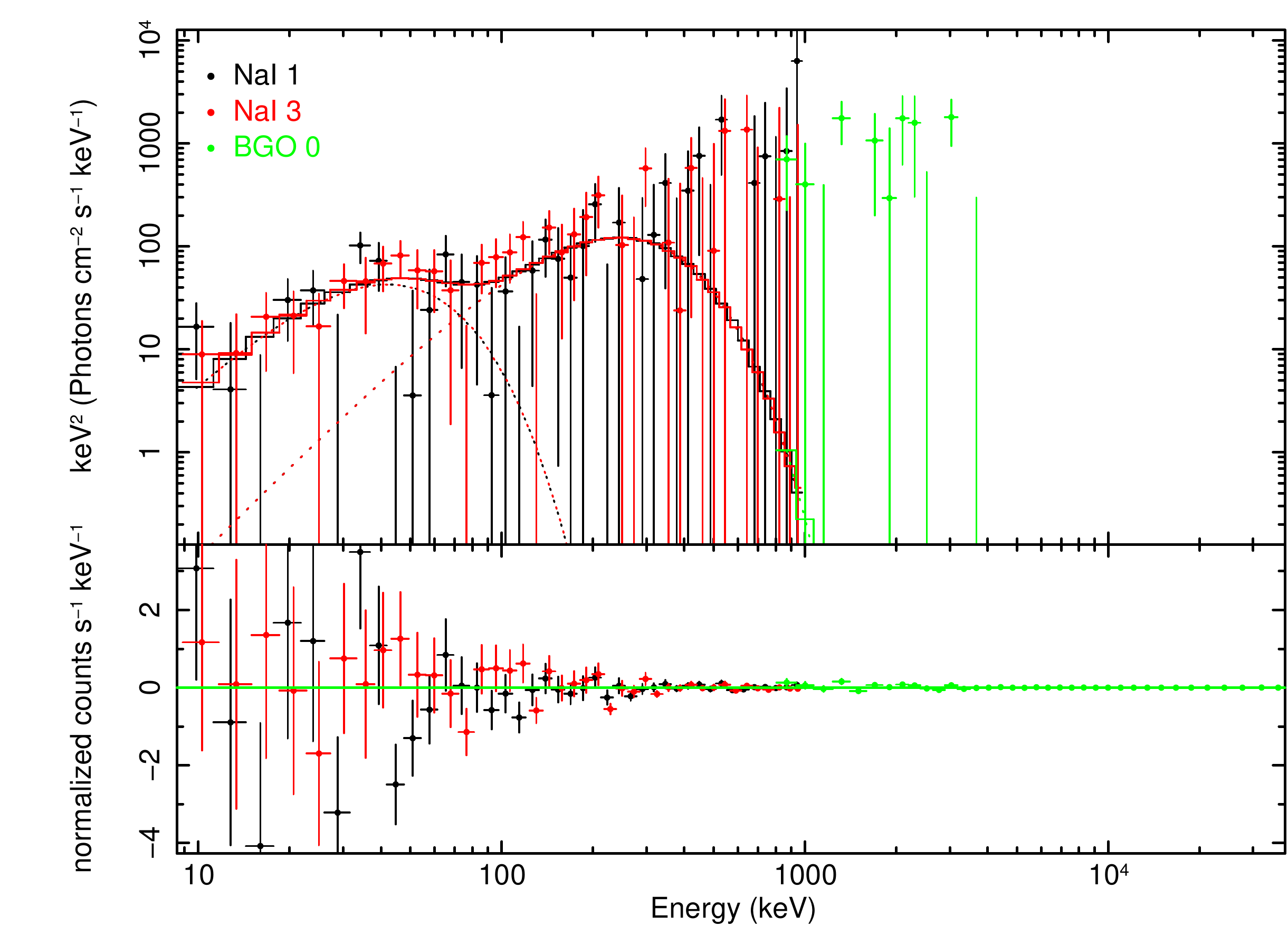}
\caption{Cluster 3 template burst, GRB091215234, fitted with two blackbodies.}
\label{fig:g3}
\end{figure}

\begin{figure}
\includegraphics[angle=270,width=\columnwidth]{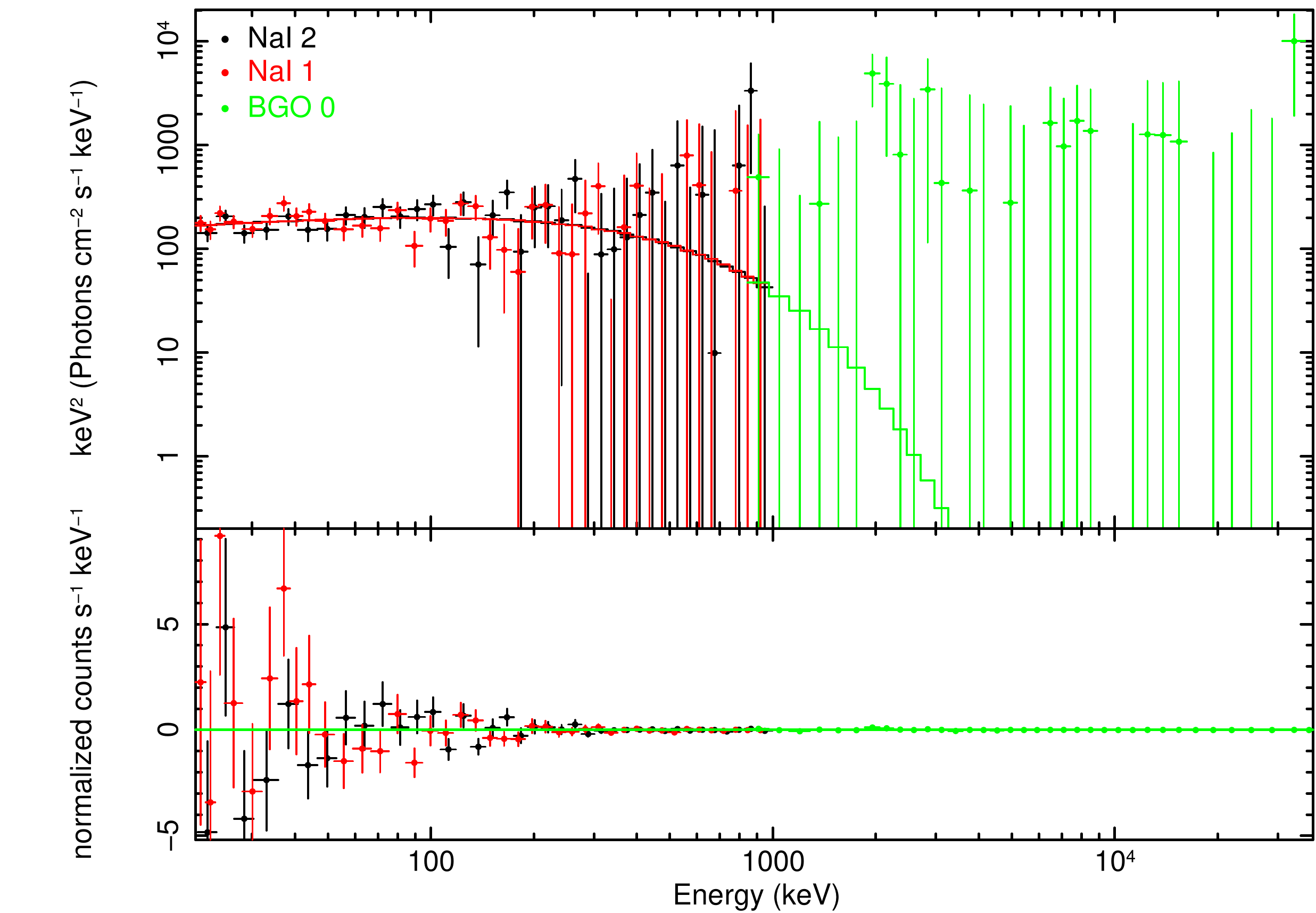}
\caption{Cluster 4 template burst, GRB100517072, fitted with the Band function.}
\label{fig:g4}
\end{figure}

\subsubsection{Cluster 1}
From the examination of its bulk Band parameter values given in the previous section, Cluster 1 can be identified with a non-thermal spectral appearance (Figure \ref{fig:g1}). Indeed, in the spectral analysis of the template bursts for this cluster, it is seen that the PGStat value has the largest reduction for the Band function. This cluster is characterized by a single smoothly broken powerlaw which is narrower compared to Clusters 2 and 4 which are also best described with this model. The $E_{pk}$ has a modest value generally around a few hundred keVs and the fluences are also modest, compared to Cluster 2. Cluster 1 averages the second highest peak flux and a high isotropic energy ($E_{iso}$) output (refer to Table \ref{tab:2}). 

\subsubsection{Cluster 2}

Cluster 2 has softer low energy index values than Group 1 for the catalogue Band fits, which reflects itself in our spectral analysis as well (Figure \ref{fig:g2}). A significant amount of improvement in the fit is obtained when these spectra are described with a Band function, as compared to the other models tried. This cluster is also described by a single break and a wide spectrum and it contains the brightest bursts in our sample which is evident from its high $E_{iso}$ and peak flux average values.

\subsubsection{Cluster 3}
The bursts in Cluster 3 are very rich in spectral features, which clearly distinguishes them from the rest of the sample (Figure \ref{fig:g3}). The best fit model is perceived to be two blackbodies from the comparison of PGStat values with an occasional need for an additional powerlaw component. The main spectral shape that is being captured with two blackbodies is a double break spectrum with a non-flat feature in between the two breaks. This kind of a spectral feature in between the lower and higher energy breaks is what gives these spectra a quite "wiggly" appearance. Cluster 5 also contains the second hardest $\alpha$ values which are captured by the blackbody fits. Members of this cluster tend to be quite faint with low peak flux averages.

\subsubsection{Cluster 4}
Cluster 4 is found to be best described with a Band model as Groups 1 and 2 but it exhibits quite a different spectral shape with a characteristic weak break at high energies and very flat low energy power-law indices, producing much broader spectra (Figure \ref{fig:g4}). There are occasional fluctuations at energies lower than 20 keV, which could be interpreted as a break, however features at these energies are affected by the lower effective area of the instrument and hence, analysis only including higher energies are preferred for this cluster. The group distinguishes itself from Clusters 1 and 2 by its very low peak flux averages as well and the $E_{iso}$ is also low for this cluster.

\subsubsection{Cluster 5}

\begin{figure}
 \includegraphics[angle=270, width=\columnwidth]{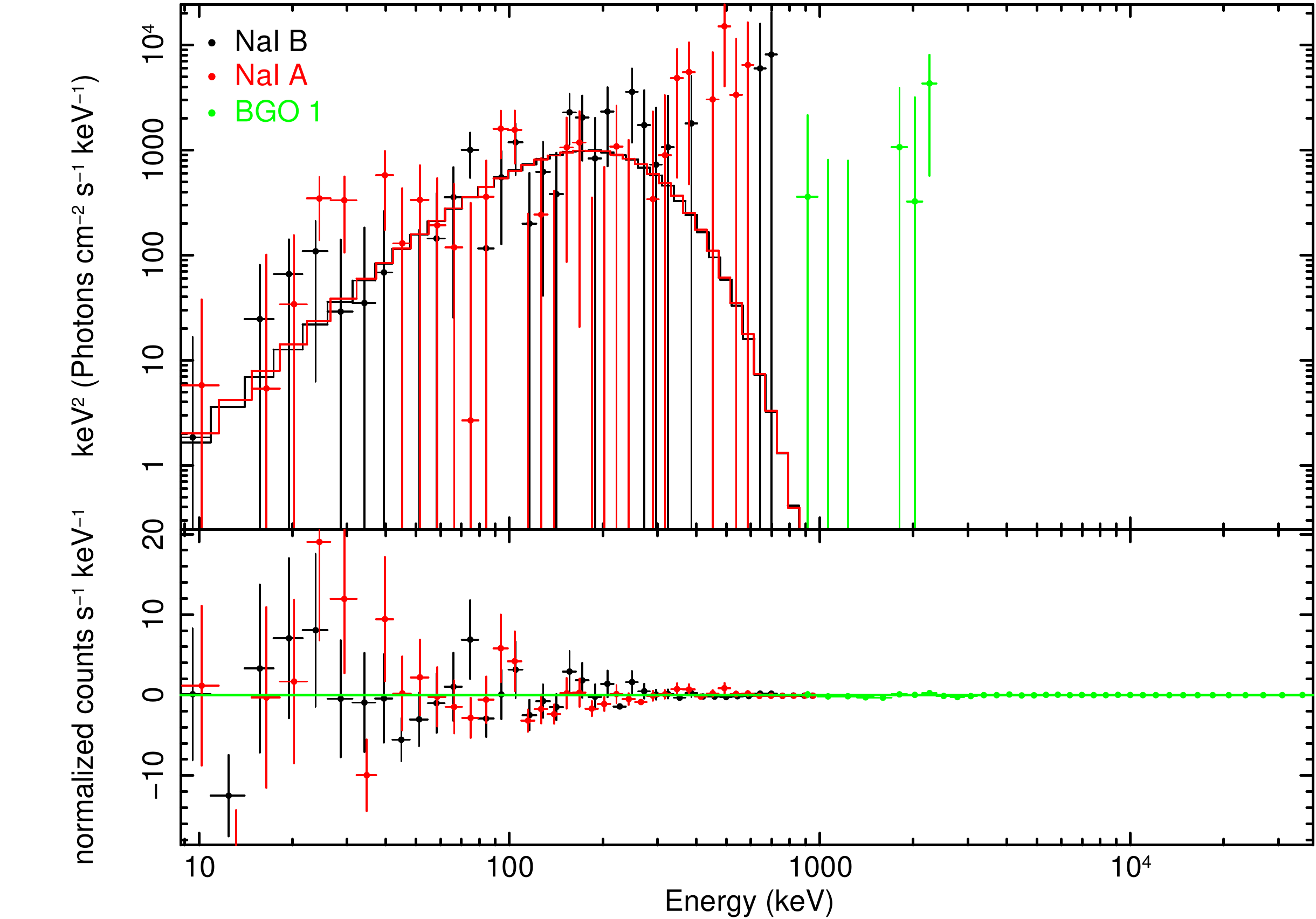}
\caption{Cluster 5 template burst, GRB100805300, fitted with a single blackbody.}
\label{fig:g5}
\end{figure}

Among our clusters, Cluster 5 has the most extreme properties concerning the low energy power law index, $T_{\rm 90}$ as well as the minimum variability time scale. This is strongly reflected in the spectral analysis of the 10 most probable bursts in this cluster, which can be well described  by a single narrow spectral component with a very steep low energy index, namely a blackbody or a broadened blackbody that can be captured by a smoothly broken power-law such as the Band function or a multicolor blackbody (Figure \ref{fig:g5}). We note, however, that these 10 top probability bursts, with a median of 0.25, are harder than the full sample, which has a median around 0. Many bursts in this cluster therefore have $\alpha <$ 0 which should be taken into account when considering the general properties of cluster 5 (see Section \ref{sec: 4} for more details). There is no information on $E_{iso}$ due to the lack of measured redshifts for this cluster, although its peak flux averages are comparable to that of clusters with brighter bursts.

\subsection{Temporal morphology of each group}
\label{sec:33}

GRB light-curves are very various in their appearances. This seemingly chaotic behavior is possibly stemming from very many distinct processes contributing to the shaping of the light-curves that are observed. Dentangling each one of them is deemed to be a daunting process, however a useful one in understanding the spectral properties better. Here, we attempt to at least classify some of the very abundant lightcurve behaviors depending on their temporal characteristics. We start this task by defining a smoothness parameter (S) for the temporal morphology of a GRB,
\begin{equation}
S = \frac{\Delta t_{min} / T90}{\sigma x, t_{min}}
\label{eq:1}
\end{equation}
where $\Delta t_{min}$ is the minimum time scale variability and $\sigma x, t_{min}$ is the fractional flux variation level at $\Delta t_{min}$ as investigated in ~\cite{Golkhou&Butler2015}.

Defined this way, the smoothness parameter is able to capture the appearance of time variability over the range of $T_{90}$, taking into account how significant the fluctuations during $\Delta t_{min}$ is by taking into account $\sigma x, t_{min}$. A low value of the smoothness parameter indicates a light-curve with many peaks or one heavily variable peak that may look damped. Larger smoothness parameters indicate a burst with a single, smooth pulse that can be interpreted with a simpler template. For a decent assessment of the light-curves with this measure, a model independent peak flux cut is required, which we determined as the peak flux on 64 ms timescale being larger than 5 $\rm{photons} \,\, cm^{-2} s^{-1}$ as a minimum. Bursts that fall below this flux cut are occasionally too low in counts and hence, give little understanding of how these three parameters work to produce each distinct light curve. This is why both fluence values used in the clustering and peak flux values are given in Tables \ref{tab:1} and \ref{tab:3}. Fluences are used in the clustering to be able to capture the full energetics of the bursts and for the convenience of being able to compare the results with the previous studies which used fluences. It is worth mentioning however that we have verified the robustness of the clustering results by doing the clustering for peak fluxes as well.

The variability measures are given in Table \ref{tab:2} and conclusions are drawn in \S \ref{sec:disc_dt}.

\subsection{Redshift distribution and fluence biases}
\label{sec:redshift}

Most of the bursts used in the clustering analysis do not have measured redshift. For the 27 bursts (9, 12, 5, 1, and 0 in each cluster, respectively) that have values we can assess the median redshift for the different clusters (see Table \ref{tab:2}). However, since the number is small any firm conclusion cannot be drawn.

The lack of knowledge of the redshift will propagate into an added dispersion of measured quantities of $E_{\rm pk}$, $T_{\rm {90}}$, and fluence. \citet{Perley2016} find the redshift distribution of GRBs to mainly be in the range of $0.6<z<4.0$, peaking in the range $1.5<z<2.5$ (see also, e.g., \citet{LeMehta2017}). The observed values of $E^{\rm obs}_{\rm pk} = E' (1+z)^{-1}$, where $E'$ is the peak energy in the progenitor rest frame, which thus translates into a dispersion of a factor of {3 (factor of 1.4 for the peak in the $z$-distribution)}. This is much smaller than the observed dispersion of $E^{\rm obs}_{\rm pk}$ (seen in, e.g., Fig. \ref{fig:FEpk}). For the duration, one would expect $T_{90} = T'_{90}(1+z)^{1-a}$, where $a\sim 0.5$ is the intrinsic dependence of the duration on the photon energy \citep{Fenimore1995,LeeP1997}, leading to a dispersion of 1.4 (1.8 for the bursts in the distribution peak). Indeed,  \citet{KP2013} concluded that any time dilation effect is masked by intrinsic and instrumental effects.
Finally, for the fluence ${\cal{F}} = \epsilon (1+z) / 4 \pi \, d^2_{\rm L}$, where $\epsilon$ is the total energy and $d_{\rm L}$ is the luminosity distance \citep{PetrosianLee1996,Meszaros2011} . This leads to a dispersion of a factor of 33 (2.5 for the peak range), which again is smaller than the observed dispersion (Fig. \ref{fig:FT90}). However, this dispersion is of the size of the dispersions of the individual clusters. Indeed, all the effects of the unknown redshifts will lead to a fuzziness in any clustering of bursts properties, most noticable in the value of the fluence.

We also point out that bursts that are intrinsically weak, for instance due to large redshift or due large viewing angles, might only be partly detected by the GBM.  If a significant fraction of the burst emission is lower than the instrument background level, then only part of the duration and the fluence will be measured. This could introduce a bias towards short and low fluence bursts. This should, however,  only affect the weakest bursts, that are  close to the detection threshold of the instrument \citep{KP2013}. In addition, the dispersions in fluence and $T_{\rm 90}$ of strong bursts are large, as can be seen in  Figure 4. Assuming that these dispersions reflect the intrinsic dispersions, also valid for the weakest bursts, then the effects of such a biases are not expected to be dominant.

 \section{Discussion}
  \label{sec: 4}

The main question that we want to answer is if there is evidence for clustering of burst properties and if this can give hints to whether there is a single or multiple emission processes involved. 

There are a number of effects that is expected to smear our any distinct clusters. First, we rely on Band function fits. In some cases, the underlying physical spectrum might be different. In such cases the Band parameters are only a proxy of the actual shape. Second, as mentioned in \S \ref{sec:redshift}, the unknown redshift will add additional dispersion. Third, it should also be noted that since we are only considering the spectrum of the light curve peak, any evolution of the type of spectrum is not captured. For instance, the ratio of the thermal and non-thermal component can vary through out a burst (discussed below in \S \ref{sec:BBBand}). It can also be imagined that a purely thermal burst transitions into a synchrotron burst and vice versa (e.g. \citet{Guiriec2013}, Zhang et al. arXiv:1612.03089). 
Due to all these effects, we therefore only expect marginal evidence of clustering from the parameters distributions alone. We note, however, that in a similar study on BATSE bursts, five cluster were also identified based on fluence, duration and spectral information, supporting the statistical result for the clustering \citep{chattopadhyay2017}. 

If the clustering, that we determine, is due to different emission processes involved, the properties of the different clusters should reflect the particularities of the individual emission processes. By studying the cluster properties from the point-of-view of the emission process can thus  lend further support of both the existence, as well as  the  cause of the clustering.  

To address this question further, we therefore discuss below the $\alpha$-distributions and the time-variability of the five clusters identified.

 \subsection{The $\alpha$-distribution and emission mechanism}
 \label{sec:51}
 
As mentioned in Section \ref{sec: 3}, the main parameters for the clustering are $E_{\rm pk}$, $T_{90}$ and fluence (see figure 3). It is therefore noteworthy that the five clusters do have different $\alpha$-distributions despite of the fact that $\alpha$ had little impact in identifying the clusters. 

Figure \ref{fig:violin} shows the $\alpha$-distribution for the five groups and we note that clusters 2 and 4 are conspicuous. First, they have exclusive contributions of $\alpha < 0$, and therefore are consistent of being purely non-thermal. At the same time the width of the $\alpha$-distributions are narrower and more symmetric than the other groups. 
Second, we note that these two clusters comprise the longest bursts, that can be seen by, for instance, from their median $T_{90}$-values. Third, neglecting cluster 5, which contain the short bursts, 
cluster 2 and 4 have large $\Epk$-values.

It is therefore suggestive that the peaks and the width of the $\alpha$-distributions of cluster 2 and 4 closely match the predictions made for GBM observations of synchrotron emission \citep{Burgess2014a}. They showed that the observed distribution should have narrow peaks at $-0.8$ and $-1.5$. The narrowness for clusters 2 and 4 is remarkable, taking into account the measurement errors on $\alpha$ as well as the possibility of bursts being wrongly assigned to each cluster, which would all increase the width of the distribution. Therefore, it can easily be argued that these two clusters are dominated by bursts that are due to synchrotron emission. Moreover, since synchrotron emission is expected to give a broader dispersion in $\Epk$ compared to photospheric emission, the observed broadness of the $\Epk$ distribution supports this interpretation for above mentioned clusters. The reason to expect a broader $\Epk$ distribution is that, for synchrotron, $\Epk^{synch} \propto \gamma_{\rm el}^2 \Gamma B_\perp$, where $\gamma_{\rm el}$ is the typical electron Lorentz factor, $\Gamma$ is the bulk Lorentz factor, and  $B$ is the typical magnetic field strength. Variations in any of these parameters will naturally cause a dispersion. On  the other hand, photospheric emission is expected to have quite narrow peak energy distributions \citep{beloborodov2013, Vurm2016}.

\begin{figure}
\includegraphics[angle=0, width=\columnwidth]{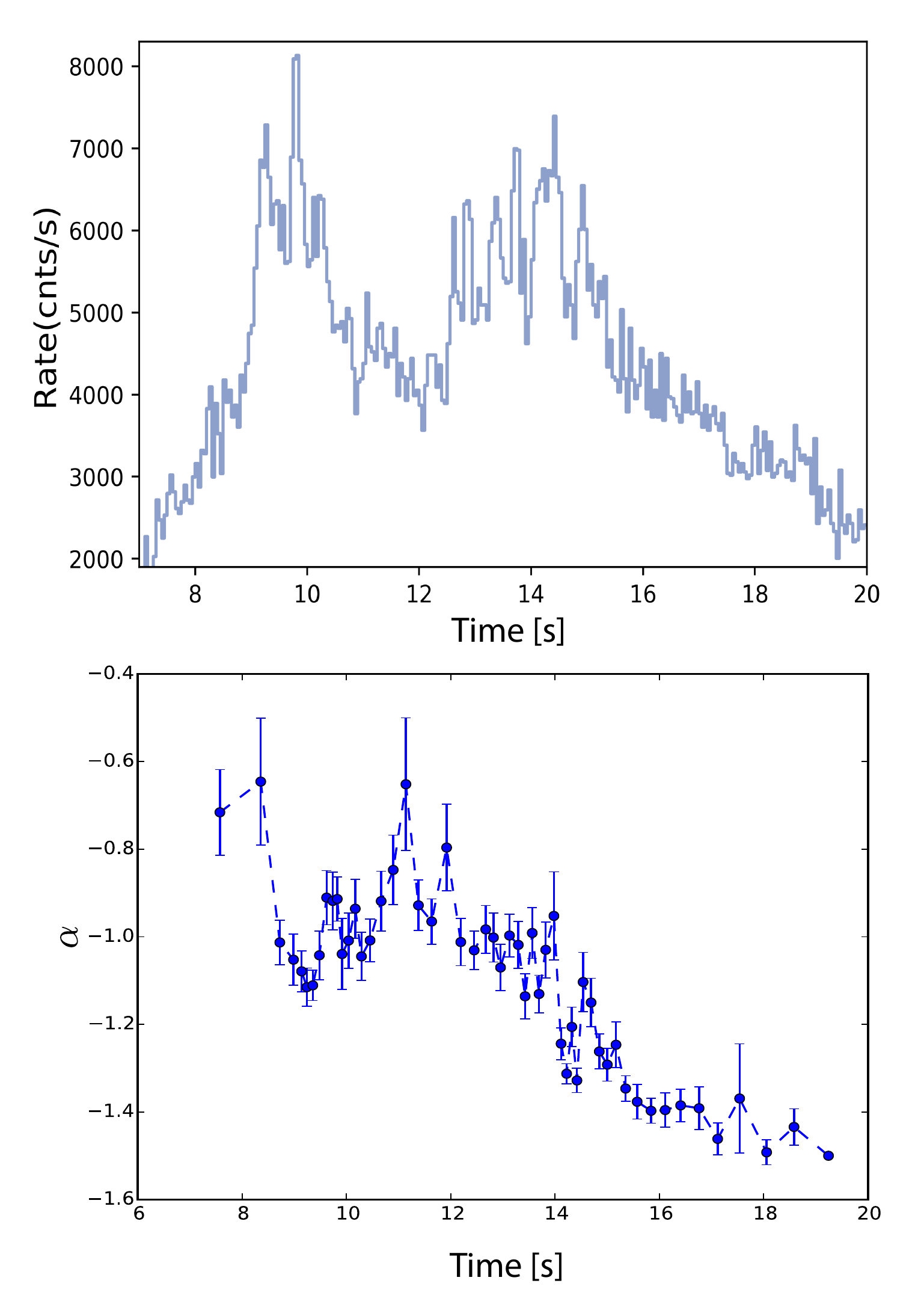}
\caption{The light curve and $\alpha$-evolution for the synchrotron bursts GRB130606.  The $\alpha$ varies within the expected values for synchrotron emission.}
\label{fig:burst1}
\end{figure}

\begin{figure}
\includegraphics[angle=0, width=\columnwidth]{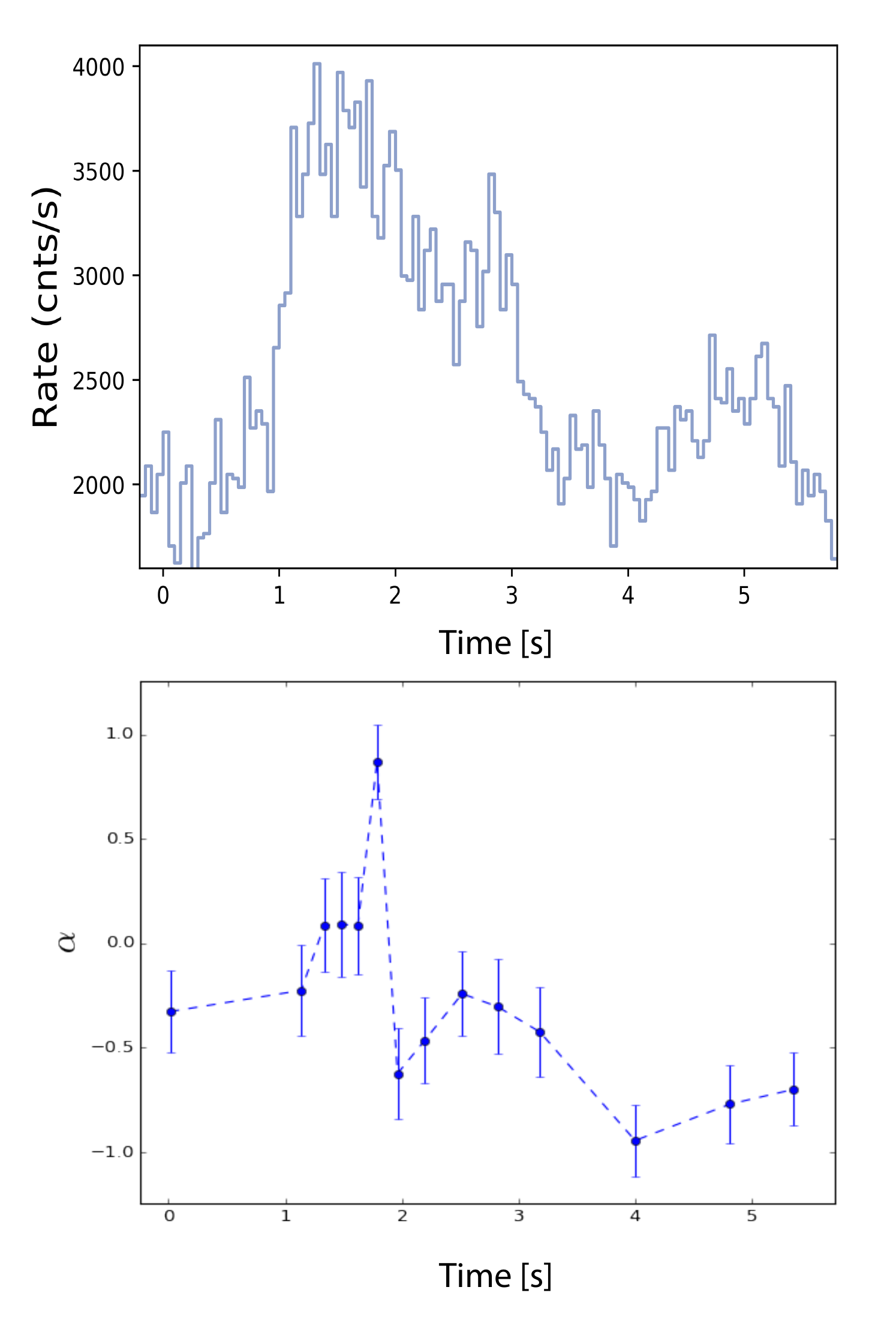}
\caption{The light curve and $\alpha$-evolution for the synchrotron bursts GRB130220. The $\alpha$ varies within the expected values for photospheric emission. }
\label{fig:burst2}
\end{figure}

It is important to bear in mind that in the clustering analysis above we only use one time-resolved measurement of $\alpha$ per burst: The peak-flux value. In order to properly assess the emission mechanism during a burst obtaining the full spectral evolution is desirable. To further examine the behaviour of bursts in cluster 2 (the most populated), we therefore examined the time-resolved evolution of $\alpha$ within the most probable bursts within the cluster. We find that $\alpha$ stays within the expected range for synchrotron emission.  We illustrate this on GRB130606 which is assigned to cluster 2 with high significance and has a $\alpha = -0.68$ at the light curve peak (this is the value used for the clustering study). Figure \ref{fig:burst1} shows the $\alpha$ and light curve of this burst. The main point to note here is that the spectra varies over the burst, there is spectral evolution. We note that the burst is consistent with synchrotron emission throughout the evolution, but the $\alpha$ value changes from $\sim -0.7$ to $\sim -1.5$, that is, from the slow to fast cooling regimes. 
Another way of assessing the range of $\alpha$-values that occur during a burst is to study the  time-integrated spectrum. The averaged value of the {\it time-integrated} $\alpha$-value for the 5 most probable bursts in cluster 2 is $<\alpha> = -1.002$, which is well within the allowed range for synchrotron emission.

We will therefore denote this group {\it the synchrotron group}, since it is dominated by bursts with properties that are consistent with synchrotron emission. The group consists of bright, long bursts, with a single spectral peak, and contains 37 \% of the bursts in the analysed sample.

The three remaining clusters (1, 3, and 5), on the other hand, do still have a significant fraction of bursts that have $\alpha > 0$. It can therefore be argued that many of the bursts must have been produced through a photospheric mechanism, such as subphotospheric dissipation. One could, however, also imagine that these groups contain bursts from a mixture of emission mechanisms, some synchrotron and some photospheric. However, among the bursts that have been assigned to clusters 1, 3, and 5 with a  probability larger than 0.8, only 8.5\% are consistent with $\alpha < -0.8$. The vast majority of the bursts are thus inconsistent with synchrotron emission. We further note that the parameter distributions of the fluence, $E_{\rm pk}$, $\beta$, and $T_{90}$ do not vary for bursts with $\alpha$ below and above  $\alpha = -0.8$. This property further supports the single emission interpretation for the majority of these bursts. We also performed analysis of the time-resolved spectral evolution within these bursts. This is illustrated by the example in Figure \ref{fig:burst2}, which shows time-resolved data for GRB130220. The peak flux value (used in the clustering analysis) for this case is $\alpha = -0.32$.  During the bursts $\alpha$ evolves strongly, but is limited in the range from -1 to 1. Similarly, the average $<\alpha>$-value for the time-integrated spectra of bursts in cluster 1 can be investigated. For the 5 most probable bursts in cluster 1 the averaged value is $<\alpha> = -0.586$. This value is significantly much harder than the corresponding value found for cluster 2, given above ($-1.002$).

From a theoretical point-of-view many spectral shapes can be produced by the photosphere \citep[e.g.][]{Peer2006, Vurm2016}.  The diversity of spectral types that can be fitted by photospheric models is illustrated by GRB090618, which was successfully fitted by \citet{Ahlgren15}, while the Band function fits yield $\alpha$-values in the range -0.8 to -1 \citep{Izzo2012}. Another example is GRB100724B which was fitted by two different photospheric models \citep{Ahlgren15,Vianello2017}, while the  $\alpha$-values lie in the range of -0.75 to -0.45 \citep{Guiriec2011,Vianello2017}. This shows that a large dispersion in $\alpha$-values can be reproduced by photospheric models.

Based on these arguments, we therefore suggest that these clusters should form a {\it photospheric group} since they are dominated by bursts that are inconsistent with synchrotron emission,  but consistent with emission from the photosphere. The group comprises clusters 1 (dimmer, single-peaked bursts), 3  (multi-break bursts) and 5 (mainly short bursts). These group contains a majority of all bursts, namely 63\%.

\subsection{Photospheric versus synchrotron bursts}

Based on the arguments in the previous section, we suggest that approximately 1/3 of all bursts are caused by synchrotron emission while approximately 2/3 of bursts are due to emission from the photosphere.

\begin{figure}
\includegraphics[angle=0, width=\columnwidth]{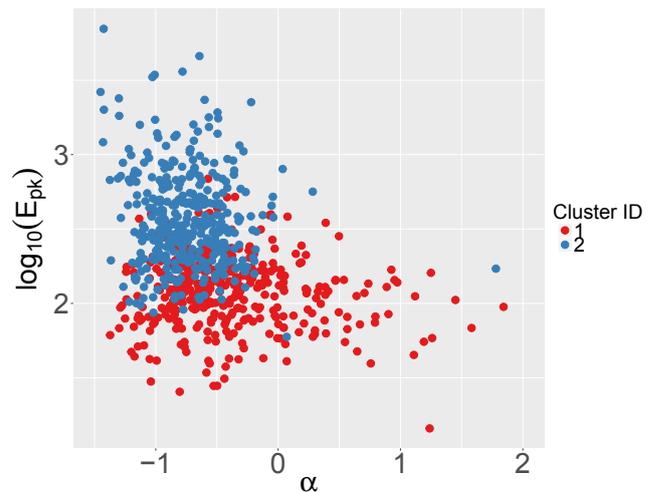}
\caption{Peak energy ($E_{\rm pk}$) versus $\alpha$ for cluster 1 (SPD bursts) and 2 (synchrotron bursts). The $\alpha$ distribution is cut at 2.}
\label{fig:EA}
\end{figure}

\begin{figure}
\includegraphics[angle=0, width=\columnwidth]{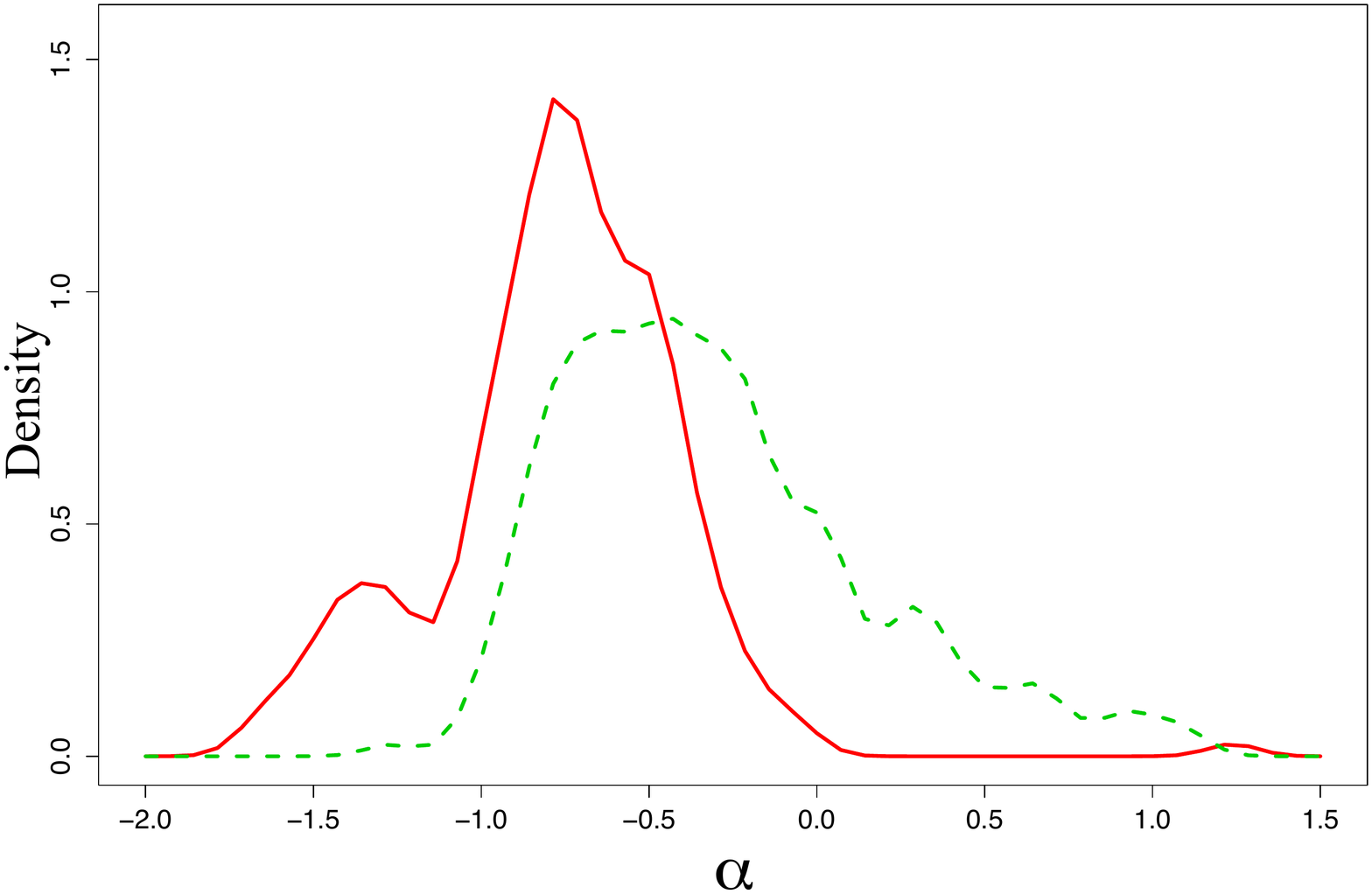}
\caption{Density distributions of $\alpha$ for two samples: the combination of cluster groups 2 and 4 (red curve), and the combination of cluster groups 1,3 and 5 (green curve). The green curve represents bursts that have spectra that are interpreted to have a photospheric origin while the red curve  represents bursts that have spectra that are consistent with synchrotron emission.}
\label{fig:AD}
\end{figure}

The $\alpha$-distributions between the synchrotron and the photospheric groups are significantly different. The two-sample Kolmogorov-Smirnov test gives $D = 0.47$, while the Wilcoxon rank sum test with continuity correction gives W = 22, which both significantly reject the hypothesis that the two distributions are drawn from the same distribution.  Figure \ref{fig:EA} shows the two distributions in the $\alpha$-$E_{\rm pk}$ plane, in which a clearly separation can be seen. To further illustrate this, in figure \ref{fig:AD} the density distribution of $\alpha$-values for bursts in these two groups are shown. For this plot we have selected bursts in our sample which have $\alpha <3$ and $\Delta \alpha < 1.0$, and that have a probability of at least 0.8 to belong to a cluster. This was done in order to only select the bursts with well determined $\alpha$, and group assignments, since we want to focus on bursts which clearly reveal their emission spectra. 
The synchrotron bursts are plotted with the red curve and the photospheric bursts are plotted with the green, dashed curve.

It is clear from figure  \ref{fig:AD} that the synchrotron bursts have two narrow peaks at the expected values. Approximately 20\% of the synchrotron bursts are in the fast cooling peak.
In contrast, the photospheric bursts do not have a preferred $\alpha$-value, but rather have a very broad peak. We note that there is a small a peak at $\alpha \sim 0.3$, which is close to the expected value of coasting phase photosphere, which should have an asymptotic value of $\alpha \sim 0.4$ \citep{Beloborodov2010}.  Moreover, the photospheric $\alpha$-distribution appears to be limited by $\alpha =$ -1 and $= 1$. The latter distribution is indeed what is expected for subphotospheric dissipation emission for low magnetised outflows \citep{Peer2006, Vurm2016}.  
The hard spectral limit of 1 is the Rayleigh-Jeans' value while the soft spectral limit can be assigned to the value that is expected from Comptonisation of a soft synchrotron emission contribution between the Wien zone (at $\tau \sim 100$) and the photosphere (at $\tau \sim 1$). Indeed, \citet{Vurm2016} finds that $\alpha = -1$ is a natural value to expect in a case of continuous dissipation through out the jet. A much softer $\alpha$ is only expected in cases with large magnetisation.

\begin{figure}
\includegraphics[angle=0, width=\columnwidth]{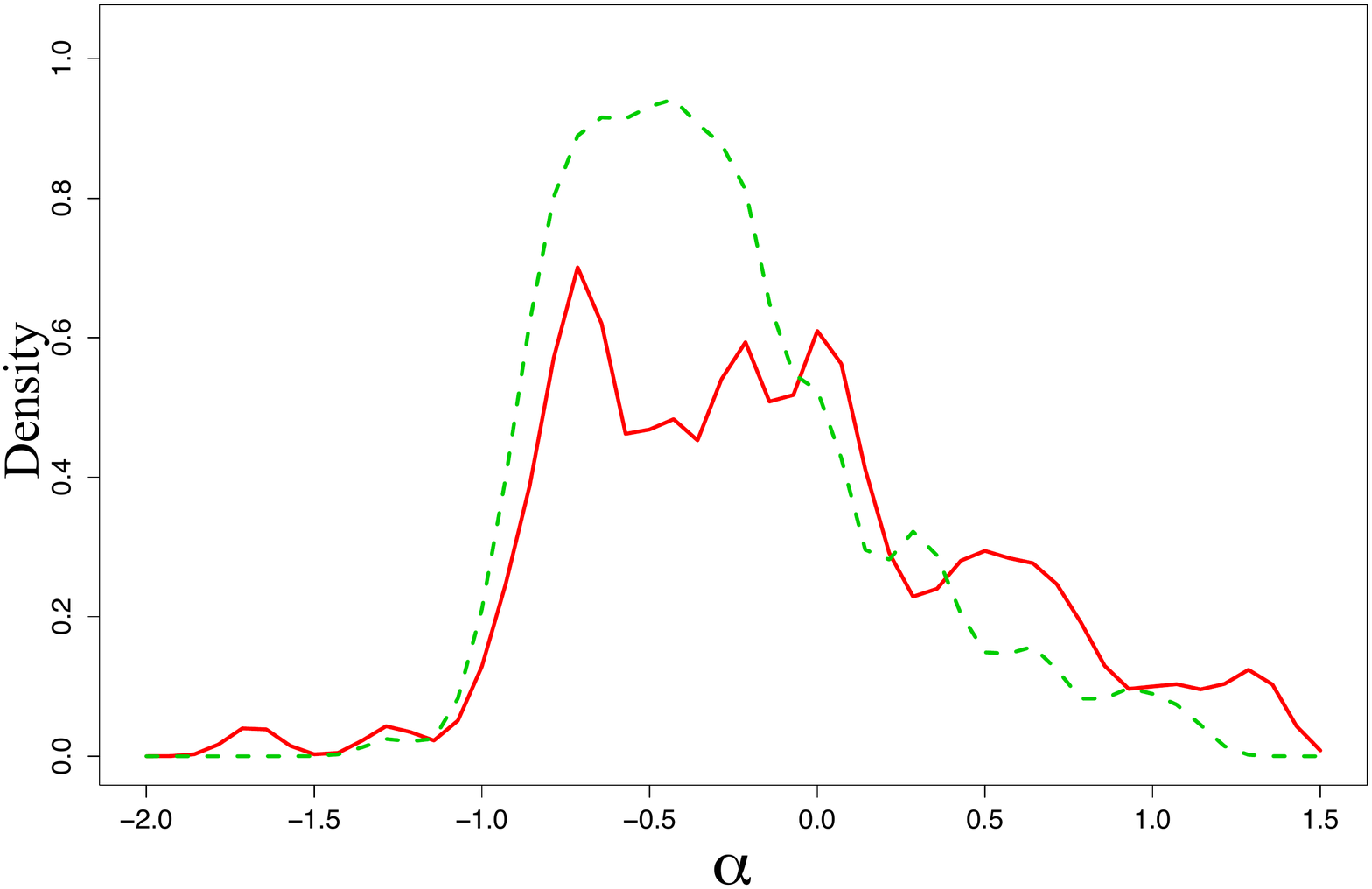}
\caption{Density distributions of $\alpha$ for two samples: (i) The photospheric group, clusters 1, 3, and 5 (green, dashed curve) and (ii) all short bursts in the sample, independent of cluster assignment (red, solid curve).}
\label{fig:AD2}
\end{figure}

In figure \ref{fig:AD2} we plot the $\alpha$-distribution for all short bursts (see also \citet{Nava2011}). We select all bursts in our sample that have $T_{\rm 90} < 2$ s. In comparison we also plot the $\alpha$-distribution of the photospheric bursts identified above (clusters 1,3, and 5). The two distributions cover the same range, which suggests that the prompt phase in most short bursts are due to photospheric emission.

\subsection{Time variability and emission mechanism}
\label{sec:disc_dt}

Further differences between the clusters are summarised in Table \ref{tab:2}. It is clear that the synchrotron bursts have more variable light curves. This is shown by the short $\Delta t_{\rm min}$ as well as the small value of the smoothness parameter $S$. This result is consistent with the finding in \citet{Dichiara2016} who found that high $\Epk$ bursts tend to be more variable on shorter time scales. In more detail, they found a strong correlation between the slope of the power density spectrum  (PDS) of the light curve and the $\Epk$ values. Since the synchrotron group are characterised by large $\Epk$ values, this correlation is consistent with short time variability scale.

Short variability time-scale poses a problem for the synchrotron interpretation for these bursts. The reason is that synchrotron spectra with $\alpha = -2/3$ are in the slow-cooling regime,
that is, most of the electrons have not had time to cool below the injection frequency. This sets strong constraints on the typical energy of the emitting electrons, $\gamma_{\rm el} \sim 10^5-10^6$ \citep{BeniaminiPiran2013}. 
This value is much higher than expected for internal shocks \citep{bosnjak_daigne_dubus2009}, which is assumed to explain highly variable lightcurves. However, large $\gamma_{\rm el}$ can be obtained for external shocks (\citet{Panaitescu1998}; \citet{Burgess2016}, \citet{duffell_macfadyen2015}). In that case the light curve variability should be low, due to the large emission radii. This is in contrast to what is suggested by the $dt_{\rm min}$ and PDS measurements. 

Another way of relaxing the condition of slow cooling and still maintaining the observed electron distribution is a marginally fast-cooling scenario \citep{Daigne2011}. 
In such a case the cooling frequency is close to the minimum injection frequency of the electrons. Indeed, as shown in Figure \ref{fig:burst1}, which shows the spectral evolution in GRB130606, the cooling regime indeed seems to vary between fast and slow cooling, which is an indication of a marginally fast-cooling scenario.  

Figure \ref{fig:burst1}, also shows that the variability of the light curve changes during the burst. It is suggestive that when $\alpha \sim - 0.7$ the light curve is less variable. The highly variable periods 9--10.4 s and 12.5--15.5 s occur during periods $\alpha < -2/3$.  Therefore, the minimum variability time scale need not occur when the spectrum is in the slow-cooling regime. This alleviates the constraints for synchrotron emission set by the time variability (see further analysis in Acuner et al. 2017, in prep.).

\subsection{Band + blackbody spectra}
\label{sec:BBBand}

One of the models tested for above consisted of a blackbody in addition to a Band spectrum. Such spectra have previously been successfully fitted to many bursts \citep[e.g.][]{Guiriec2011, Guiriec2013, Iyyani2013,  Axelsson2012, Burgess2014a, Preece2014, Nappo2017}. However,  \citet{Burgess2015b} showed that a model that combines synchrotron emission and a blackbody can at most account for little more than half of the  $\alpha$-distribution of Band function fits. 

None of the clusters have been identified as having Band + blackbody as the best fit spectra according to our method, which was limited to making detailed spectral analysis to the bursts assigned with the highest probability to a cluster. However, many bursts have been observed to have a blackbody component on top of a Band spectrum within the GBM energy band, such as GRB100724A \citep{Guiriec2011}, GRB110721A \citep{Axelsson2012},  GRB081224887, GRB090719A, GRB100707A \citep{Burgess2014a}, GRB090926A and GRB080916C \citep{Guiriec2015a}, and GRB131014A \citep{Guiriec2015b}. Moreover, for instance GRB151027A has a blackbody component at $\sim 10$ keV, which thus is not detectable within the GBM energy band \citep{Nappo2017}. A striking fact is that all of these nine bursts are assigned to cluster 2, that is, the synchrotron cluster{\footnote{  We consistently study and interpret time-resolved data since only then the physical nature of the emission can directly be assessed. We note that there are a few burst for which a Band+BB model has been fitted to the time-integrated spectrum. However, interpretation of such cases must be done with caution since deviations from a Band spectrum could be an artifact of spectral evolution during the integrated period that is studied \citep{Burgess2015a}.}. Concluding from this, a fraction of the bursts in cluster 2 can still have a subdominant signature of the photosphere, in form of a blackbody. Further exploration should be made to identify how large this fraction is. A consequence of the existence of such bursts in cluster 2 is that it makes photospheric emission identified in an even larger fraction of bursts (in addition to clusters 1, 3 and 5). Furthermore, assuming that the non-thermal component is synchrotron emission, a subdominant blackbody component would increase the expected deviation of the measured $\alpha$ from a synchrotron value, if the spectrum is fitted with a Band function only \citep[see, e.g.,][]{Guiriec2015a, Burgess2015b}. This would further increase the width of the $\alpha$-distribution.

\subsection{What determines the emission mechanism in GRBs?}

What determines if a burst is dominated by synchrotron or photospheric emission? Three possibilities are given (i) by the jet dissipation pattern, (ii) by the jet content, and (iii) by the viewing angle.

(i) Strong dissipation below the photosphere would energise the photospheric emission, producing a variety of spectral shapes \citep{Rees&Meszaros2005, Peer2006, Beloborodov2010}. 
On the other hand, if the flow is initially smooth the photosphere could be weakened due to adiabatic cooling and dissipation in the optically-thin region, either as internal shocks or external shocks, would produce a strong synchrotron signal. 

(ii) Alternatively, the difference could be assigned to the jet content. If thermal energy dominates, a strong photospheric component could be expected, in particular, if dissipation occurs below the photosphere and/or if the photosphere occurs close to the saturation radius. On the other hand, if the flow is dominated by Poynting flux the photospheric component is expected to be weak and a strong synchrotron component can arise \citep{Zhang&Peer2009, Gao&Zhang2014, BeguePeer2015}.

(iii) Finally, variations in appearance of bursts could be caused by different viewing angle, i.e.,  different angles between the jet axis and the line of sight \citep[e.g][]{Ioka2001, Salafia2016}.
In particular, since the jet is expected to be surrounded by a cocoon of shocked jet material, whose properties could be similar to that of the jet \citep{Nakar2017}, the  observer could detect either the jet or the shocked jet cocoon, depending on the viewing angle. This might lead to different clusters identified above. For instance, the long, variable, synchrotron bursts (clusters 2 and 4) might be cases which are observed head-on, as the jet itself. Dissipation, such as shocks, occurring in the jet cocoon might enhance the subphotospheric dissipation leading to photospheric bursts as in clusters 1 and 3. 
If the dissipation is strong, the peak energy is expected to be larger and the $\alpha$ value to be soft \citep{Vurm2016}. 
The strength of dissipation therefore might explain the difference between cluster 1 (higher $E_{pk}$, soft $\alpha$) and cluster 3 (lower $E_{pk}$, harder $\alpha$). Moreover, it can be speculated that a situation could arise in which parts of the jet and parts of the jet cocoon are observed simultaneously, if the two parts can be distinguished within $\Gamma^{-1}$. This could then be the cause of bursts observed with blackbody on top of the Band function (see \S \ref{sec:BBBand}). Finally, in the varying-viewing-angle scenario the emitting surface of jet is only 40\% of the jet/jet-cocoon system, based on the number of observed bursts.\footnote{The photospheric clusters 1 and 2 contain 602  bursts and the synchrotron clusters 2 and 4 contain 421 bursts (Table 1)}

Nonetheless, it could still be argued that all bursts are photospheric, since this model can produce a large variety of spectral shapes, depending on the dissipation pattern in the jet.  However, such an interpretation must be able to explain the astonishing coincidence of the peaks of clusters 2 and 4 and their narrowness. Moreover, bursts with shallow sub-peak spectra ($\alpha \sim -1.5$) and  bursts with very broad peaks have not yet been fully explored in such models.

\section{Conclusions}
  \label{sec: 5}

What emission mechanism is responsible for GRBs is still an unsettled question. The distributions of measured quantities of the prompt phase emission do not naturally match neither the assumption that all bursts are due to synchrotron emission nor the assumption that all bursts are from emission from the photosphere. A possible reason for this could be that both emission from the photosphere and optically-thin synchrotron emission are operating, and dominate differently in individual bursts. In order to investigate this we performed a clustering analysis of the observed properties ($\alpha$, $\beta$, $\Epk$, $T_{\rm 90}$, and fluence) of all the bursts {\it Fermi}/GBM catalogue. We identify five clusters in a sample of 1151 bursts.

Further analysis of the parameter distributions of these individual clusters, including additional information such as time variability, spectral properties and energetics, reveal an astonishingly concordance with what is expected from pure synchrotron and photospheric emissions. Based on this analysis we therefore argue that
the main division should be made in two groups. Around 1/3 of the bursts are consistent with synchrotron emission, which forms the first group. This group contains bursts that are bright, long, single peaked spectra with large $\Epk$-values. The second group with 2/3 of all burst are consistent with photospheric emission and contains a cluster of long single peaked bursts, a cluster of multi-peaked burst, and finally, a cluster of short bursts ($T_{\rm 90} < 2$ s). 

We further argue that this division could be due to the dissipation pattern in the jet, alternatively due to whether the jet is dominated by thermal or magnetic energy, or finally due to the viewing angle. 

Finally, we point out that our analysis cannot exclude other possible explanation, such as that all bursts are due to subphotospheric dissipation, or models with a mixture of emission components. However, such models still need to give a natural explanation for the distribution of observed parameters characterising the emission.

\vskip 5mm

{\bf Acknowledgements}

{We thank Drs. Magnus Axelsson, Andrei Beloborodov, Christoffer Lundman, and Asaf Pe'er for enlightening discussions. This research has made use of data obtained through the High Energy Astrophysics Science Archive Research Center Online Service, provided by the NASA/Goddard Space Flight Center. We acknowledge support from the Swedish National Space Board and the Swedish Research Council (Vetenskapsr{\aa}det). FR is supported by the  G\"oran Gustafsson Foundation for Research in Natural Sciences and Medicine.}

\bibliographystyle{mnras}

\clearpage

\appendix
\label{sec:6}

\section{Doubly broken power law model}
\label{sec:DBPL}

Doubly broken powerlaw model is defined as,
\begin{equation}
pl = [(E_{1}/E)^{(p*\alpha)}+ (E_{1}/E)^{(p*\beta)}+[(E_{2}/E)^{(p*\delta)}.[(E_{1}/E_{2})^{(p*\alpha)} +  (E_{1}/E_{2})^{(p*\beta)}]]]^{-1/p}.
\end{equation}

\section{Lists for burst samples in 5 clusters}
\begin{table*}
\centering
	\caption{Cluster 1 sample bursts.}
\begin{tabular}{llllllllllllll}
  \hline
 \hline
    ~ & GRB ID & $T_{90}$ & $E_{pk} $ & Flux  & Fluence  & $\alpha$ & $\beta$  & $\Delta t_{min} $ & $\sigma x, t_{min}$  & S & Cluster & Probability \\ 
    ~& ~ & [s] & [keV]& [pht/cm$^2$.s] &  [erg/cm\textsuperscript{2}] & ~ & ~ & [s] & ~ & ~ &~& ~ \\  
   \hline
  1 & 090709630 & 22.27 & 113.54 & 5.11& 2.2$ \times10^{-5}$ & -0.34 & -2.96& 1.21 & 0.41 & 0.13 & 1 & 0.96 \\ 
  2 & 090804940 & 5.57 & 113.24 & 40.69 & 1.4$ \times10^{-5}$ & -0.32 & -3.48 & 0.41 & 0.33 & 0.22 & 1 & 0.97 \\ 
  3 & 091127976 & 8.7 & 55.45 & 102.97 & 2.07$\times10^{-5}$& -0.51 & -2.27 & 0.1 & 0.42 & 0.028& 1 & 0.97 \\ 
  4 & 100612726 & 8.58 & 104.53 & 28.42 & 1.4$\times10^{-5}$ & -0.46 & -3.17 & 0.39 & 0.2 & 0.23 & 1 & 0.97 \\ 
  5 & 100816026 & 2.05 & 131.92 & 19.88 & 3.7$\times10^{-6}$ & -0.13 & -3.2 & 0.33 & 0.42 & 0.39 & 1 & 0.98\\ 
  6 & 100907751 & 5.38 & 79.33 & 4.92 & 7.3$\times10^{-7}$& -0.15 & -3.28 & 3.08 & 1.84 & 0.312 & 1 & 0.98 \\ 
  7 & 101016243 & 3.84 & 152.34 & 14.41 & 2.4$\times10^{-6}$ & -0.73 & -3.34 & 0.34 & 0.62 & 0.144 & 1 & 0.96\\ 
  8 & 101216721 & 1.92 & 169.61 & 25.08 & 3.04$\times10^{-6}$ & -0.64 & -3.47 & 0.11 & 0.2 & 0.274 & 1 & 0.95 \\ 
  9 & 120122300 & 16.70 & 105.42 & 5.01 & 2.6$\times10^{-6}$ & -0.28 & -3.71 & 0.58 & 0.48 & 0.072 & 1 & 0.97 \\ 
  10 & 120427054 & 5.63 & 122.56 & 18.49 & 7.4$\times10^{-6}$ & 0.24 & -2.88 & 0.89 & 0.54 & 0.294 & 1 & 0.97 \\ 
   \hline
   
\end{tabular}
\end{table*}

\begin{table*}
\centering
	\caption{Cluster 2 sample bursts.}
    
\begin{tabular}{llllllllllllll}
  \hline
   \hline
      ~ & GRB ID & $T_{90}$ & $E_{pk} $ & Flux  & Fluence  & $\alpha$ & $\beta$  & $\Delta t_{min} $ & $\sigma x, t_{min}$  & S & Cluster & Probability \\ 
    ~& ~ & [s] & [keV]& [pht/cm$^2$.s] &  [erg/cm\textsuperscript{2}] & ~ & ~ & [s] & ~ & ~ &~& ~ \\ 
  \hline
  1 & 080817161 & 60.3 & 371.9 & 17.4 & 5.3$\times10^{-5}$ & -0.75 & -2.02 & 0.24 & 0.15 & 0.027 & 2 & 0.993 \\ 
  2 & 081215784 & 5.6 & 752.52 & 148.47 & 5.5$\times10^{-5}$ & -0.53 & -2.4 & 0.04 & 0.3 & 0.025 & 2 & 0.997 \\ 
  3 & 100414097 & 26.5 & 477.98 & 28.16 & 8.9$\times10^{-5}$ & -0.72 & -2.5 & 0.03 & 0.18 & 0.006 & 2 & 0.993 \\ 
  4 & 100724029 & 114.7 & 472.31 & 27.073 & 2.2$\times10^{-4}$ & -0.67 & -2.002 & 0.06 & 0.08 & 0.007 & 2 & 0.998 \\ 
  5 & 100826957 & 84.9 & 546.18 & 37.34 & 1.6$\times10^{-4}$ & -0.7 & -2.2 & 0.106 & 0.14 & 0.009 & 2 & 0.997 \\ 
  6 & 100918863 & 86.02 & 896.28 & 10.94 & 8.9$\times10^{-5}$ & -0.85 & -2.84 & 2.06 & 0.34 & 0.07 & 2 & 0.989 \\ 
  7 & 101014175 & 449.42 & 630.15 & 71.22 & 2$\times10^{-4}$ & -0.95 & -2.22 & 0.05 & 0.23 & 0.0005 & 2 & 0.994 \\ 
  8 & 110921912 & 17.7 & 457.4 & 41.36 & 3.6$\times10^{-5}$ & -0.63 & -2.3 & 0.024 & 0.21 & 0.007 & 2 & 0.991 \\ 
  9 & 120624933 & 271.4 & 683.86 & 21.25 & 1.9$\times10^{-4}$ & -0.79 & -2.34 & 0.13 & 0.2 & 0.003 & 2 & 0.996 \\ 
  10 & 120711115 & 44.03 & 1357.24 & 44.67 & 1.9$\times10^{-4}$ & -0.83 & -2.21 & 0.08 & 0.5 & 0.0034 & 2 & 0.999 \\ 
   \hline
\end{tabular}
\end{table*}

\begin{table*}
\centering
	\caption{Cluster 3 sample bursts.}
\begin{tabular}{llllllllllllll}
  \hline
   \hline
     ~ & GRB ID & $T_{90}$ & $E_{pk} $ & Flux  & Fluence  & $\alpha$ & $\beta$  & $\Delta t_{min} $ & $\sigma x, t_{min}$  & S & Cluster & Probability \\ 
    ~& ~ & [s] & [keV]& [pht/cm$^2$.s] &  [erg/cm\textsuperscript{2}] & ~ & ~ & [s] & ~ & ~ &~& ~ \\ 

  \hline
  1 & 091017861 & 2.6 & 37.7 & 4.3 & 4.5$\times10^{-7}$& 1.24 & -1.8 & 1.783 & 1.5 & 0.45 & 3 & 0.99 \\ 
  2 & 091026485 & 3.3 & 48.4& 4.96 & 6$\times10^{-7}$ & 0.57 & -1.76 & 1.152 & 0.78 & 0.44 & 3 & 0.99 \\ 
  3 & 091215234 & 4.4 & 51.3& 5.09 & 9.9$\times10^{-7}$ & 0.6 & -1.73 & 0.825 & 0.53 & 0.36 & 3 & 0.989\\ 
  4 & 110204179 & 28.7 & 58.7 & 5.3 & 3$\times10^{-6}$ & 0.6 & -1.6 & 0.279 & 0.28 & 0.035 & 3 & 0.974\\ 
  5 & 110411629 & 23.6 & 78.2& 7.8 & 3.6$\times10^{-6}$ & 0.19 & -1.84 & 4.331 & 1.11 & 0.17 & 3 & 0.957 \\ 
  6 & 110528624 & 69.6 & 49.9 & 4.6 & 4.6$\times10^{-6}$ & 0.9 & -1.95 & 0.739 & 0.4 & 0.027 & 3 & 0.952 \\ 
  7 & 110818860 & 67.1 & 54.9 & 4.9 & 5.2$\times10^{-6}$ & 1.25 & -1.64 & 2.791 & 0.64 & 0.065 & 3 & 0.979 \\ 
  8 & 110819665 & 16.4 & 64.3 & 18.6 & 3$\times10^{-6}$ & 0.7 & -2 & 0.249 & 0.57 & 0.027 & 3 & 0.959 \\ 
  9 & 120102416 & 20.2 & 54.2 & 5.6 & 2.6$\times10^{-6}$ & 0.005 & -1.65 & 0.597 & 0.4 & 0.074 & 3 & 0.967 \\ 
  10 & 120506128 & 2.3 & 72.03 & 4.03 & 2.9$\times10^{-7}$ & -0.46 & -2.04 &0.675 & 0.76 & 0.39 & 3 & 0.992 \\ 

\hline
\end{tabular}
\end{table*}

\begin{table*}
\centering
	\caption{Cluster 4 sample bursts.}
\begin{tabular}{llllllllllllll}
  \hline
   \hline
    ~ & GRB ID & $T_{90}$ & $E_{pk} $ & Flux  & Fluence  & $\alpha$ & $\beta$  & $\Delta t_{min} $ & $\sigma x, t_{min}$  & S & Cluster & Probability \\ 
    ~& ~ & [s] & [keV]& [pht/cm$^2$.s] &  [erg/cm\textsuperscript{2}] & ~ & ~ & [s] & ~ & ~ &~& ~ \\ 
  \hline
1 & 081206604 & 7.94 & 1211.6 & 3.02 & 5$\times10^{-7}$ & -1.7 & -3.4 & 1.86 & 0.88 & 0.27 & 4 & 0.999 \\ 
  2 & 090428552 & 31.49 & 76.8 & 9.87 & 6$\times10^{-6}$ & -1.5& -2.07 & 1.17 & 0.42 & 0.09 & 4 & 0.986 \\ 
  3 & 100201588 & 122.11 & 179.72 & 4.5 & 1$\times10^{-7}$ & -1.38 & -3.95 & 2.15 & 0.29 & 0.061 & 4 & 0.995 \\ 
  5 & 100517072 & 55.81 & 96.76 & 18.7 & 7$\times10^{-6}$ & -1.44 & -3.02 & 0.07 & 0.22 & 0.006 & 4 & 0.999 \\ 
  6 & 110426629 & 356.36 & 48.27 & 8.6 & 3$\times10^{-5}$ & -1.42 & -3.64 & 3.9 & 0.38 & 0.029 & 4 & 0.999 \\ 
  7 & 110803783 & 186.88 & 271.14 & 4.8 & 3$\times10^{-6}$ & -1.37 & -2.74 & 2.3 & 0.68 & 0.018 & 4 & 0.968\\ 
  8 & 111228657 & 99.84 & 94.72 & 27.6 & 1.8$\times10^{-5}$ & -1.37 & -2.73 & 0.085 & 0.18 & 0.005& 4 & 0.994\\ 
  9 & 120210650 & 1.34 & 516.2 & 11.07 & 6.5$\times10^{-7}$ & -1.68 & -3.15 & 0.084 & 0.48 & 0.13& 4 & 0.997\\ 
  10 & 120710100 & 131.84 & 158.04 & 6.6 & 5.3$\times10^{-6}$ & -1.6 & -1.89 & 0.67 & 0.22 & 0.023 & 4 & 0.990 \\ 
   \hline
\end{tabular}
\end{table*}

\begin{table*}
\centering
\caption{Cluster 5 sample bursts.}
\begin{tabular}{llllllllllllll}
 \hline
  \hline
    ~ &GRB ID & $T_{90}$ & $E_{pk} $ & Flux  & Fluence  & $\alpha$ & $\beta$  & $\Delta t_{min} $ & $\sigma x, t_{min}$  & S & Cluster & Probability \\ 
    ~& ~ & [s] & [keV]& [pht/cm$^2$.s] &  [erg/cm\textsuperscript{2}] & ~ & ~ & [s] & ~ & ~ &~& ~ \\ 
  \hline
  1 & 090617208 & 0.2 & 492.6 & 18.9 & 9.4$\times10^{-7}$& 0.8 & -2.18 & 0.015 & 0.54 & 0.145 & 5 & 0.999 \\ 
  2 & 100206563 & 0.18 & 566.09 & 25.37 & 7.6$\times10^{-7}$ & -0.2 & -2.38 & 0.017 & 0.37 & 0.26 & 5 & 0.999 \\ 
  3 & 100216422 & 0.2 & 509.75 & 8.99 & 3.9$\times10^{-7}$ & 0.04 & -1.72 & 0.045 & 0.78 & 0.3 & 5 & 0.999 \\ 
  4 & 100612545 & 0.6 & 505.34 & 12.32 & 2.2$\times10^{-6}$  & 0.69 & -2.09& 0.024 & 0.64 & 0.07 & 5 & 0.999 \\ 
  5 & 100625773 & 0.24 & 328.46 & 17.08 & 5.6$\times10^{-7}$  & 0.65 & -2.23 & 0.03 & 0.92 & 0.14 & 5 & 0.999 \\ 
  6 & 100805300 & 0.06 & 205.76 & 21.5 & 2.04$\times10^{-7}$  & 1.84 & -3.27 & 0.042 & 0.9 & 0.75 & 5 & 0.999 \\ 
  7 & 101026034 & 0.26 & 123.33 & 13.98 & 9.3$\times10^{-7}$  & 0.66 & -1.7 & 0.017 & 0.56 & 0.12 & 5 & 0.999 \\ 
  8 & 101204343 & 0.13 & 365.32 & 6.57 & 2.8$\times10^{-7}$  & -0.34 & -1.95 & 0.015 & 0.71 & 0.17 & 5 & 0.999 \\ 
  9 & 110409179 & 0.13 & 265.48 & 9.65 & 3.3$\times10^{-7}$  & 0.59 & -2.23 & 0.009 & 0.61 & 0.12 &5 & 0.999 \\ 
  10 & 110717180 & 0.112 & 260.98 & 18.5 & 2.5$\times10^{-7}$ & 0.03 & -1.95 & 0.01 & 0.64 & 0.15 & 5 & 0.999 \\ 
   \hline
\end{tabular}
\end{table*}

\end{document}